\documentclass[twocolumn]{aastex63}

\usepackage{array}
\usepackage{amssymb}
\usepackage{amsmath}
\usepackage[caption=false]{subfig}

\hypersetup{colorlinks}

\newcommand\Tstrut{\rule{0pt}{3ex}}       
\newcommand\Bstrut{\rule[-1.5ex]{0pt}{0pt}} 
\newcommand\TBstrut{\Tstrut\Bstrut}

\def\PSR{\textrm{PSR J0537-6910}}
\def\PSRs{\textrm{PSR J0537-6910 }}
\def\Jpsr{\textrm{J0537-6910}}
\def\Jpsrs{\textrm{J0537-6910 }}

\def\Fstat{\F{\textrm{-statistic}}}

\def\LIGO{\textrm{LIGO}}

\newcommand{\Hz}{\mathrm{Hz}}
\newcommand{\HzS}{\mathrm{Hz/s}}
\newcommand{\HzSsq}{\mathrm{Hz/s}^2}

\def\dnu{\dot{\nu}}
\def\ddnu{\ddot{\nu}}
\def\ddnuLT{\ddot{\nu}_\textrm{LT}}
\def\nLT{n_\textrm{LT}}
\def\ddnuig{\ddot{\nu}_\textrm{ig}}
\def\nig{n_\textrm{ig}}

\def\tauCW{T_\textrm{span}}
\def\tobs{t_\textrm{obs}}
\def\nuobs{\nu|_{t_\textrm{obs}}}

\def\nuK{\nu_\textrm{K}}

\def\Tst{t^\textrm{gw}_0} 

\def\Tcoh{\tauCW}
\def\Tdata{T_{\textrm{\mbox{{data}}}}}
\def\Tref{t_{\textrm{\mbox{{ref}}}}}

\def\ds{\textrm{days}}

\def\Neff{N^\textrm{eff}}

\def\fmin{f_\min}
\def\fmax{f_\max}
\def\dfmin{\dot{f}_\min}
\def\dfmax{\dot{f}_\max}
\def\ddfmin{\ddot{f}_\min}
\def\ddfmax{\ddot{f}_\max}
\def\fBand{\Delta f}
\def\dfBand{\Delta \dot{f}}
\def\ddfBand{\Delta \ddot{f}}

\def\chisq{\chi^2}
\def\chisqdegr{\chi^2_4}
\def\Fthr{2\F_\textrm{thr}}

\def\hsd{h_0^\textrm{sd}}
\def\alphaUL{\alpha^{90\%}}
\def\hUL{h_0^{90\%}}
\def\hUUL{h_0^{95\%}}

\def\Izz{I_{zz}}

\def\dnuovnu{{{|\dnu|}\over \nu}}

\def\Msun{M_\odot}

\def\min{\textrm{\mbox{min}}}
\def\max{\textrm{\mbox{max}}}

\def\dF{\delta f}
\def\ddF{\delta \dot{f}}
\def\dddF{\delta \ddot{f}}

\newcommand{\ee}[1]{\!\times\!10^{#1}}

\newcommand{\Freq}{f}
\newcommand{\fdot}{{\dot{\Freq}}}
\newcommand{\fddot}{\ddot{\Freq}}

\newcommand{\M}{\mathcal{M}}

\newcommand{\Gauss}{\mathrm{\MakeUppercase{G}}}
\newcommand{\Signal}{{\mathrm{\MakeUppercase{S}}}}
\newcommand{\Line}{{\mathrm{\MakeUppercase{L}}}}

\providecommand{\sc}[1]{\widehat{#1}}
\renewcommand{\sc}[1]{\widehat{#1}}

\newcommand{\Bayes}{\hat{B}}

\newcommand{\BSGL}{\Bayes_{\Signal\Gauss\Line}}

\newcommand{\F}{\mathcal{F}}		

\newcommand{\scFtho}{\F_*^{(0)}}

\newcommand{\sft}{{\mathrm{SFT}}}

\newcommand{\Nsft}{N_\sft}

\newcommand{\RA}{\ensuremath{{84.4476}}} 
\newcommand{\DEC}{\ensuremath{{-69.17219}}} 
\newcommand{\nLTnom}{\ensuremath{{-1.22}}} 
\newcommand{\nLTunc}{\ensuremath{{-1.22 \pm 0.04}}} 
\newcommand{\ddnuLTnom}{\ensuremath{{-7.7\ee{-22}}}} 
\newcommand{\ddnuLTunc}{\ensuremath{{-7.7\ee{-22} \pm 3\ee{-23}}}} 
\newcommand{\nignom}{\ensuremath{{7.4}}} 
\newcommand{\nigunc}{\ensuremath{{7.4 \pm 0.7}}} 
\newcommand{\ddnuigmin}{\ensuremath{{4.89\ee{-21} \pm 7\ee{-23}}}} 
\newcommand{\ddnuigmax}{\ensuremath{{2.13\ee{-20} \pm 7\ee{-22}}}} 
\newcommand{\numinOOne}{\ensuremath{{61.9344(10)}}} 
\newcommand{\numaxOOne}{\ensuremath{{61.9365(9)}}} 
\newcommand{\numinOTwoOne}{\ensuremath{{61.9261(14)}}} 
\newcommand{\numaxOTwoOne}{\ensuremath{{61.9294(11)}}} 
\newcommand{\numinOTwoTwo}{\ensuremath{{61.9236(15)}}} 
\newcommand{\numaxOTwoTwo}{\ensuremath{{61.9266(12)}}} 
\newcommand{\nudotmin}{\ensuremath{{-4.22(27)\ee{-10}}}} 
\newcommand{\nudotmax}{\ensuremath{{-2.02(11)\ee{-10}}}} 
\newcommand{\nuddotmin}{\ensuremath{{4.89(7)\ee{-21}}}} 
\newcommand{\nuddotmax}{\ensuremath{{2.13(7)\ee{-20}}}} 
\newcommand{\fminOOne}{\ensuremath{{85.9078(14)}}} 
\newcommand{\fmaxOOne}{\ensuremath{{97.2403(13)}}} 
\newcommand{\fbandOOne}{\ensuremath{{11.3325(19)}}} 
\newcommand{\fminOTwoOne}{\ensuremath{{85.8964(19)}}} 
\newcommand{\fmaxOTwoOne}{\ensuremath{{97.2291(17)}}} 
\newcommand{\fbandOTwoOne}{\ensuremath{{11.3327(25)}}} 
\newcommand{\fminOTwoTwo}{\ensuremath{{85.8929(20)}}} 
\newcommand{\fmaxOTwoTwo}{\ensuremath{{97.2247(19)}}} 
\newcommand{\fbandOTwoTwo}{\ensuremath{{11.3318(28)}}} 
\newcommand{\fdotmin}{\ensuremath{{-6.63(43)\ee{-10}}}} 
\newcommand{\fdotmax}{\ensuremath{{-2.79(15)\ee{-10}}}} 
\newcommand{\fdotband}{\ensuremath{{3.84(45)\ee{-10}}}} 
\newcommand{\fddotmin}{\ensuremath{{4.82(7)\ee{-21}}}} 
\newcommand{\fddotmax}{\ensuremath{{3.34(11)\ee{-20}}}} 
\newcommand{\fddotband}{\ensuremath{{2.86(11)\ee{-20}}}} 
\newcommand{\TstOOne}{\ensuremath{{1126623625}}} 
\newcommand{\TrefOOne}{\ensuremath{{1131937856}}} 
\newcommand{\tauOOne}{\ensuremath{{123.0}}} 
\newcommand{\TdataOOne}{\ensuremath{{6287}}} 
\newcommand{\dFOOne}{\ensuremath{{3.05}\ee{-08}}} 
\newcommand{\ddFOOne}{\ensuremath{{2.22}\ee{-14}}} 
\newcommand{\dddFOOne}{\ensuremath{{9.89}\ee{-21}}} 
\newcommand{\lgNeffOOne}{\ensuremath{{13.3}}} 
\newcommand{\FstarOOne}{\ensuremath{{34.2}}} 
\newcommand{\TstOTwoOne}{\ensuremath{{1167983370}}} 
\newcommand{\TrefOTwoOne}{\ensuremath{{1170799164}}} 
\newcommand{\tauOTwoOne}{\ensuremath{{65.2}}} 
\newcommand{\TdataOTwoOne}{\ensuremath{{4107}}} 
\newcommand{\dFOTwoOne}{\ensuremath{{5.77}\ee{-08}}} 
\newcommand{\ddFOTwoOne}{\ensuremath{{7.92}\ee{-14}}} 
\newcommand{\dddFOTwoOne}{\ensuremath{{6.65}\ee{-20}}} 
\newcommand{\lgNeffOTwoOne}{\ensuremath{{12.0}}} 
\newcommand{\FstarOTwoOne}{\ensuremath{{31.1}}} 
\newcommand{\TstOTwoTwo}{\ensuremath{{1180975619}}} 
\newcommand{\TrefOTwoTwo}{\ensuremath{{1184354596}}} 
\newcommand{\tauOTwoTwo}{\ensuremath{{78.2}}} 
\newcommand{\TdataOTwoTwo}{\ensuremath{{4790}}} 
\newcommand{\dFOTwoTwo}{\ensuremath{{4.8}\ee{-08}}} 
\newcommand{\ddFOTwoTwo}{\ensuremath{{5.5}\ee{-14}}} 
\newcommand{\dddFOTwoTwo}{\ensuremath{{3.85}\ee{-20}}} 
\newcommand{\lgNeffOTwoTwo}{\ensuremath{{12.2}}} 
\newcommand{\FstarOTwoTwo}{\ensuremath{{31.6}}}

\usepackage{multirow}

\usepackage{footmisc}

\shorttitle{First search for r-mode gravitational waves from $\PSR$}
\shortauthors{L. Fesik and M. A. Papa}

\begin{document}

\title{First search for r-mode gravitational waves from $\PSR$}

\email{liudmila.fesik@aei.mpg.de}

\author{Liudmila Fesik}
\affiliation{Max Planck Institute for Gravitational Physics (Albert Einstein Institute), Callinstrasse 38, 30167 Hannover, Germany}
\affiliation{Leibniz Universit\"at Hannover, D-30167 Hannover, Germany}

\author{Maria Alessandra Papa}
\email{maria.alessandra.papa@aei.mpg.de}
\affiliation{Max Planck Institute for Gravitational Physics (Albert Einstein Institute), Callinstrasse 38, 30167 Hannover, Germany}
\affiliation{Leibniz Universit\"at Hannover, D-30167 Hannover, Germany}
\affiliation{University of Wisconsin Milwaukee, 3135 N Maryland Ave, Milwaukee, WI 53211, USA}

\begin{abstract}

We report results of the first search to date for continuous gravitational waves from unstable r-modes from the pulsar $\PSR$. 
We use data from the first two observing runs of the Advanced LIGO network. We find no significant signal candidate and set upper limits on the amplitude of gravitational-wave signals, which are within an order of magnitude of the spin-down values. We highlight the importance of having timing information at the time of the gravitational-wave observations, i.e. rotation frequency and frequency-derivative values, and glitch-occurrence times, such as those that a NICER \citep{nicer} campaign could provide. 

\end{abstract}


\section{Introduction} \label{sec:intro}

Fast-spinning neutron stars are among the most promising sources of gravitational radiation. Non-axisymmetric deformations and ``wobbles" of rotating stars will produce quasi-monochromatic long-lasting gravitational emission -- continuous gravitational waves (CWs).
In addition, gravitational radiation can destabilize \textit{$r$-modes} -- quasi-normal stellar oscillations of rotating stars \citep{Andersson:1997xt, Friedman:1997uh,Owen:1998xg} -- and give rise to substantial continuous gravitational-wave emission. This instability is particularly interesting in hot young neutron stars because it could provide an effective spin-down mechanism \citep{Lindblom:1998wf}.

If neutron stars form in collapse processes, from the conservation of angular momentum one might expect their initial spin to be close to the theoretical maximum that their structure could support, between 500 and 2000 Hz, depending on the equation of state. The observations, however, indicate that young neutron stars present rather smaller spins. Gravitational-wave driven r-mode instabilities have been put forward as a mechanism to explain the missing young fast-rotating pulsars \citep{Andersson:1998ze} and concrete detection strategies have been proposed \citep{Owen:2010ng,Caride:2019hcv}. 

The fastest and the most energetic known young pulsar is {$\PSR$}. This object is associated with the supernova remnant N157B in the Large Magellanic Cloud, its age is estimated to be 4000 yr, and it is spinning at about $62\,\Hz$ \citep{Marshall:1998sp, Townsley_2006}. This spin frequency may be just below the predicted final frequency for the r-mode emission mechanism \citep{Alford:2012yn,Alford:2014pxa}.

The analysis of 13 years (1999--2011) of X-ray spin timing observations of \PSRs with the Rossi X-ray Timing Explorer (RXTE) has revealed an extreme glitch activity with abrupt spin-ups (\textit{glitches}) every few months and a subsequent post-glitch relaxation phase \citep{Antonopoulou:2017hwa}. A recent study of the post-relaxation phase data has found an intriguing indication: the average braking index during these periods is $\approx 7.4$ \citep{Andersson:2017fow}.

Why is this intriguing? The braking index $n$ is commonly used to describe the spin evolution of neutron stars, $\dnu(t) \propto {\nu (t)}^n$, with 
$n = \nu \ddnu/\dnu^2$, with $\nu$ indicating the spin frequency. If the star's spin evolution is driven by magnetic dipole emission $n=3$; if quadrupolar gravitational-wave emission is the culprit then $n=5$; for r-modes $n=7$, under some approximations \citep{Kokkotas:2015gea,Alford:2012yn}. This means that the measured value of the braking index $n \sim 7.4$ might suggest that unstable r-mode emission is the main driver of the spin evolution of \PSR.

With this background we perform a directed search for continuous gravitational waves from $\PSR$ assuming that the emission stems from unstable r-modes. We use data from the Advanced LIGO network (aLIGO) \citep{Vallisneri_2015,O1data,O2data} spanning the period between Sept, 2015 and Aug, 2017. Since electromagnetic (EM) observations of J0537 are not available for this period, the pulsar's rotational parameters are not precisely known and its glitch activity is unknown. 

The paper is organized as follows. We summarise relevant results from the timing analysis of EM data in Sec. \ref{sec:evolutionPSR}. Relations between the expected gravitational-wave frequency and the spin of the pulsar are defined in Sec. \ref{sec:rmode} and the search is detailed in Sec. \ref{sec:search}. Results are presented in Sec. \ref{sec:results} and discussed in Sec. \ref{sec:conclusions}.

\section{Spin evolution of $\Jpsr$}
\label{sec:evolutionPSR}

$\PSR$ is spinning at  $\approx 62\,\Hz$ with a strong spin-down rate $\dnu \approx -2 \ee{-10}\,\HzS$. Its spin evolution is usually described as the superposition of two trends: the long-term (LT) evolution that describes the spin evolution on the timescale of years and the short-term (ST) evolution that describes the post-glitch recovery phase and is appropriate for weeks after a glitch \citep{Antonopoulou:2017hwa}.

The long-term braking index $\nLT$ of $\Jpsr$ is derived by fitting the measurements of $\dot{\nu}$ at the mid-time epochs between two subsequent glitches over 13 years of data. The result is $\nLT = \nLTnom$ with a negative second-order frequency derivative $\ddnuLT = \ddnuLTnom\,\HzSsq$ \citep{Antonopoulou:2017hwa}. 

We will refer to the periods between two successive glitches as glitch-free or inter-glitch intervals. The inter-glitch spin evolution is estimated through a phase-coherent timing analysis of the $45$ known inter-glitch intervals and yields a wide range of braking indices, with most $\nig >10$ \citep{Antonopoulou:2017hwa}. A detailed analysis \citep{Andersson:2017fow} shows that the largest contributions to $\nig$ come from epochs $\le 50\,\ds$ after a glitch, indicating the existence of an early fast relaxation phase. In contrast, an asymptotic value of $\nig$ for the longer time intervals is $\approx \nignom$ (Fig. 3 in \citep{Andersson:2017fow}), which might reflect the fact that gravitational-wave emission due to r-mode instability is causing the observed spin down. Incorporating the effect of temperature and frequency on the saturation amplitude yields values of the braking index different from $\sim7$ \citep{Kokkotas:2015gea,Alford:2012yn}. In this search we assume a constant value of the braking index derived from the timing analyses and neglect these effects. For simplicity we will assume that r-mode emission sets in 50 days after a glitch. 

\begin{figure}[h!tb]
   \includegraphics[width=\columnwidth]{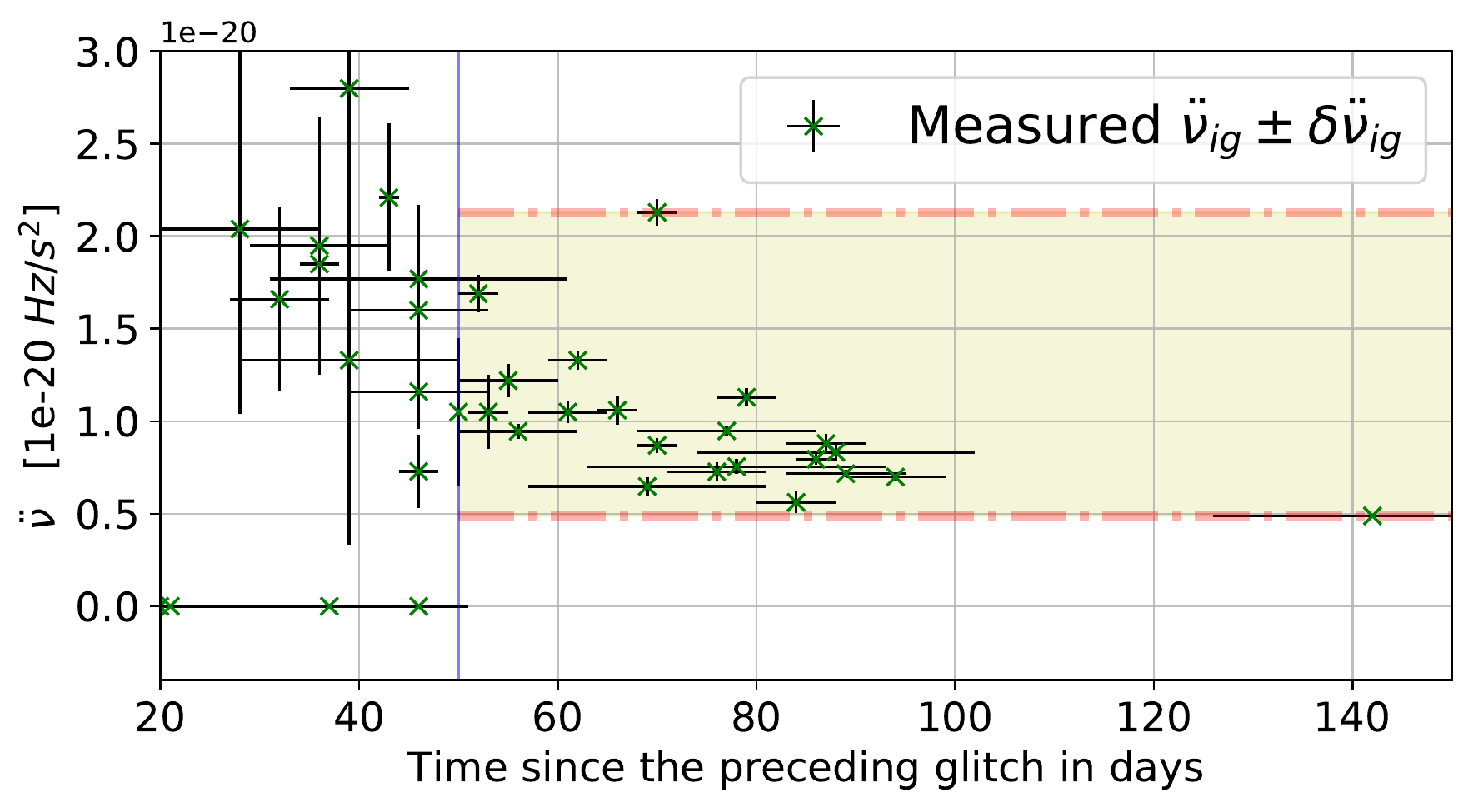}
    \caption{Distribution of measured $\ddnuig\pm \delta\ddnuig$ for every known inter-glitch period, as a function of the fit epoch. The fit epoch is 0 at the time of each glitch. (We used the data from Tab. 1 of \citep{Antonopoulou:2017hwa}.) } 
    \label{fig:nu2dot_J0537}. 
\end{figure}

\begin{deluxetable}{lcrc}[tbh!]
\tablecaption{Sky position and spin evolution parameters for $\PSR$ \citep{Townsley_2006, Antonopoulou:2017hwa}\label{tab:RotEvolution}}
\tablehead{
	\multicolumn{4}{c}{\TBstrut \textbf{Sky position}}
} 
\startdata 
\multicolumn{4}{c}{\TBstrut $\alpha \,$ [deg] $\RA$ \hspace{10pt} $\delta $ [deg] $\DEC$ } \\ \hline
\multicolumn{4}{c}{\TBstrut \textbf{Long-term evolution}} \\ \hline
\TBstrut $\nLT$   &   $\nLTunc$  & \TBstrut  $\ddnuLT\,[\HzSsq]$ & $\ddnuLTunc$ \\ \hline
\multicolumn{4}{c}{\TBstrut \textbf{Short-term evolution}} \\ \hline
\TBstrut $\nig$ & $\nigunc$  & $\ddnuig^\min \,[\HzSsq]$ & $\ddnuigmin$ \\ \hline
\TBstrut  &   & $\ddnuig^\max \,[\HzSsq]$ & $\ddnuigmax$ \\ \hline
\multicolumn{4}{c}{\TBstrut \textbf{The last observation}} \\ \hline
\multicolumn{3}{r}{\TBstrut fit epoch $\tobs\,$ [GPS]} & $1004659215$ \\ \hline
\TBstrut  & & \TBstrut $\nuobs \,[\Hz]$ &  $61.961105096 \pm 5\times 10^{-9} $ \\
\enddata  
\end{deluxetable} 

Fig.~\ref{fig:nu2dot_J0537} shows the second-order frequency-derivative values for the various inter-glitch periods as a function of the epoch of the measurement. For epochs that more than $50\,\ds$ after the glitch, $\ddnuig \in [\nuddotmin, \nuddotmax]\, \HzSsq$, as shown in Tab. 1 of \citep{Antonopoulou:2017hwa}. 

The most important values from the timing analysis of $\PSR$ are summarized in Tab. \ref{tab:RotEvolution}. 

\newpage
\section{Gravitational-wave emission from r-modes}
\label{sec:rmode}
The strongest gravitational waves are expected from the quadrupole ($l=m=2$) r-mode, so we concentrate on this. The gravitational-wave frequency $f$ associated with this mode depends on the neutron star structure and its rotation frequency $\nu$ in a non-trivial manner \citep{Owen:1998xg,Lindblom:1998wf}. 
We follow the prescription of \citep{Caride:2019hcv} and use the following relations: 
\begin{equation}
	\label{eq:CWfreq_rmode}
	\begin{cases}
    		f/\nu = A - B \, ( \nu/ \nuK)^2  \\ 
		\fdot/\dnu = A - 3 B \, ( \nu/ \nuK)^2 \\
		\ddot{f}/\ddnu = A - 3 B \, ( \nu/ \nuK)^2  {\left( 1-{\frac{2}{n}}\right)}\\	
	\end{cases}
\end{equation}
with $n$ being the braking index during the r-mode phase, $\nuK$ the Kepler frequency of the star, and the quantities $1.39 \leq A \leq 1.57$ and $0 \leq B \leq 0.195$ encoding information on the neutron star structure. Based on the observed highest spin frequency of pulsars at $716$ Hz, following \citep{Caride:2019hcv}, we take 506 Hz as a lower bound for $\nuK$. Using this value yields a broader (i.e. more conservative) search range than for a $\nuK$ in line with the standard estimates for neutron stars and higher by a factor of 2 or 3 \citep{PhysRevD.46.4161,Paschalidis:2016vmz}.

The uncertainties in the values of $A$ and $B$ give rise to ranges of values for the gravitational-wave frequency and frequency derivatives. Since  $A$ is always $\gg 3 B ( \nu/ \nuK)^2 $ they take the form

\vspace{-.8em}
\begin{equation}
\label{eq:searchRanges0}
\begin{cases}
 \vspace{.4em}
     \left( 1.39 - 0.195~{ {\nu^2}\over {\nuK^2}}~ \right) \nu \leq  f \leq 1.57 ~\nu \\
   \vspace{.4em}
   {   \left( 1.39 - 0.585~{ {\nu^2}\over {\nuK^2}}~ \right) |\dnu|  \leq  |\fdot| \leq 1.57 ~|\dnu|     } \\
   \vspace{.4em}
   { \left( 1.39 - 0.585~{ {\nu^2}\over {\nuK^2}}~ \right){\left( 1-{2\over 7}\right)} ~\ddnu \leq  \ddot{f}  \leq 1.57 ~\ddnu   }
    \end{cases}
\end{equation}
and we note that $\dot\nu=-|\dot\nu|$ and $\fdot=-|\fdot|$. 
\vspace{1em}

\section{The Gravitational-wave Search}
\label{sec:search}

\begin{figure}[h!tb]
   \includegraphics[width=\columnwidth]{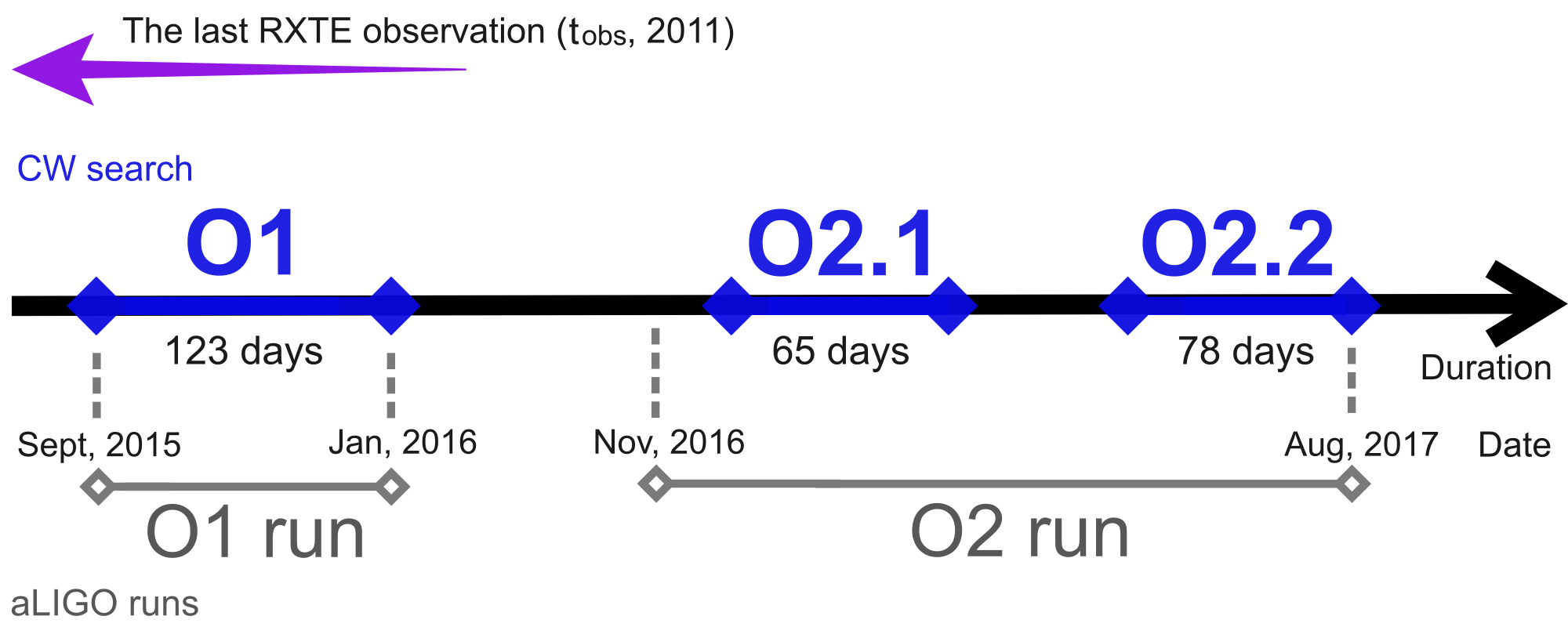}
    \caption{The data from the O1 and O2 aLIGO observing runs used for our searches.} 
    \label{fig:data}
\end{figure}

\subsection{The data input}
\label{subsec:DataInput}

We use the data from the two Advanced $\LIGO$ (aLIGO) detectors located in Hanford (WA) and Livingston (LA), USA \citep{Abbott:2007kv}. We search the publicly available data from the first two observing runs O1 and O2 \citep{O1data,O2data,Vallisneri_2015}. O1 took place between September 12, 2015 and January 19, 2016 and covered about $4$ months of data \citep{TheLIGOScientific:2016agk,TheLIGOScientific:2014jea}. The second run (O2) operated from 2016 November 30 to 2017 August 25 and includes a significant gap in the data between 2017 March 15th and June 8th \citep{LIGOScientific:2018mvr,o2datasummary}. This gap naturally provides two stretches of contiguous data, which we consider as two independent time baselines for our searches: \textit{O2.1} and \textit{O2.2}, as shown in Fig. \ref{fig:data}. 

The search input consists of short time baseline Fourier transforms (SFTs) \citep{SFTs}, from data segments $1800\,$s long. Instrumental and environmental spectral disturbances are removed to avoid contamination of the results, as done in previous searches \citep{Covas:2018oik,Abbott:2017pqa}. 

\subsection{The signal waveform}
\label{sec:waveform}

The signal at each detector, $h(t)$, is a superposition of the two polarization waveforms $h_+ (t)$ and $h_\times(t)$
\begin{equation}
	h(t)=F_+ (\alpha,\delta,\psi ;t) h_+ (t) + F_\times (\alpha,\delta,\psi; t) h_\times(t),
	\label{eq:signal}
\end{equation}
where  $F_+ (\alpha,\delta,\psi;t)$ and $F_\times(\alpha,\delta,\psi;t)$ are the detector beam pattern functions and
\begin{eqnarray}
	h_+ (t)  =  A_+ \cos \Phi(t) \nonumber \\
	h_\times (t)  =  A_\times \sin \Phi(t).
	\label{eq:monochromatic}
\end{eqnarray}
If $\iota$ is the angle between the total angular momentum of the star and the direction from the star to Earth
\begin{eqnarray}
	A_+  & = & {1\over 2} h_0 (1+\cos^2\iota) \nonumber \\
	A_\times & = &  h_0  \cos\iota. 
	\label{eq:amplitudes}
\end{eqnarray}
\noindent $h_0$ is the intrinsic gravitational-wave amplitude, $(\alpha,\delta)$ are the
right-ascension and declination for the source, and $\psi$ is the
orientation of the wave-frame with respect to the detector frame. Due
to Earth's motion, the orientation between the detector and the source is changing
all the time, which makes $F_{+,\times}$ time-varying.
$\Phi(t)$ is the phase of the gravitational-wave signal at time
$t$. If with $\tau$ we indicate the arrival time of the wave with
phase $\Phi(t)$ at the solar system barycenter (SSB), then 
\begin{multline}
\label{eq:phiSSB}
\Phi(\tau) = \Phi_0 + 2\pi [ f (\tau-{\tau_\textrm{ref}})  +
\\ {1\over 2} \dot{f} (\tau-{\tau_\textrm{ref}})^2 + {1\over 6} \ddot{f} (\tau-{\tau_\textrm{ref}})^3 + \cdots ].
\end{multline}
If the frequency derivatives are non-zero, the reference time $\tau_\textrm{ref}$ (or $\Tref$) determines the frequency scale. The transformation between detector time $t$ and solar system barycenter time
$\tau$ is
\begin{equation}
 \label{eq:tau(t)} \tau(t) = t+ \frac{\mathbf{r}(t) \cdot \mathbf{n}}{c} +
\Delta_{\mathrm{E}\odot} - \Delta_{\mathrm{S}\odot}\,,
\end{equation} 
where $\mathbf{r}(t)$ is the position vector of the detector in the SSB
frame, $\mathbf{n}$ is the unit vector pointing to the source,
and $c$ is the speed of light; $\Delta_{\mathrm{E}\odot}$ and $\Delta_{\mathrm{S}\odot}$
are respectively the relativistic Einstein and Shapiro time
delays.  We refer the reader to \citep{Jaranowski:1998qm} for further details.

\subsection{The signal-parameter ranges}
\label{sec:ParamSpace}

The position of \PSRs is known with high accuracy, so if we knew precisely its spin frequency $\nu$ and its derivatives $\dot\nu, \ddot\nu$, from Eq.~(\ref{eq:searchRanges0}) we could determine the range of possible values of the r-mode gravitational-wave frequency $f$ and its derivatives $\fdot,\fddot$. However, since there are no timing data available for the O1 and O2 data period \citep{Antonopoulou:2017hwa}, $\nu, \dot\nu,\ddot\nu$ are not known precisely. 

For each search we set the reference time $\Tref$ in the middle of each observation period. At that reference time we determine the range of values for $\nu\in[\nu_\min, \nu_\max]$, $\dot\nu\in[\dot\nu_\min, \dot\nu_\max]$ and $\ddot\nu\in[\ddot\nu_\min, \ddot\nu_\max]$. From Eq.~\ref{eq:searchRanges0} the corresponding gravitational-wave frequency and derivatives ranges are then 
\begin{equation}
	\label{eq:searchRanges}
	\begin{cases}
  	\vspace{.4em}
    	\left( 1.39 - 0.195~{ {\nu^2}\over {\nuK^2}}~ \right)\bigg\rvert_{\nu_\min} \nu_\min \leq  f \leq 1.57 ~\nu_\max \\
   \vspace{.4em}
   {  \left( 1.39 - 0.585~{ {\nu^2}\over {\nuK^2}}~ \right)\bigg\rvert_{\nu_\max}  |\dot\nu|_\min \leq  |\fdot| \leq 1.57 ~|\dot\nu|_\max    } \\
   \vspace{.4em}
   {\left( 0.9929 - 0.4179~{ {\nu^2}\over {\nuK^2}}~ \right)\bigg\rvert_{\nu_\max} \ddot\nu_\min   \leq  \ddot{f}  \leq 1.57 ~\ddot{\nu}_\max  }.
    \end{cases}
\end{equation}

\begin{figure}[h!tb]
   \includegraphics[width=\columnwidth]{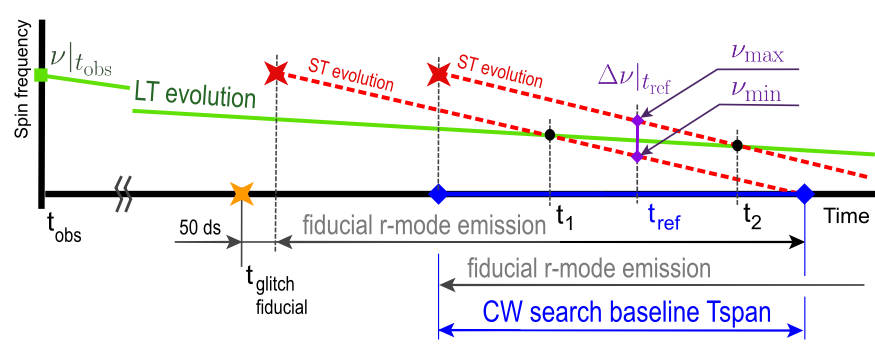}
    \caption{Spin evolution of the $\PSR$ under the two different scenarios discussed in the text.} 
    \label{fig:main_scheme}
\end{figure}

The range of possible values for $\nu$ and $\dot\nu$ stems from
our ignorance on when glitches occurred, bracketing the gravitational-wave observations. If we assume that there is r-mode emission throughout our observation periods, then the spin state of the star at each $\Tref$ only depends on how long before $\Tref$, the previous glitch happened. We consider two extremes: 1) r-mode emission sets-in just at the beginning of our observation period and lasts for a very long time; 2) r-mode emission sets-in a long time before the beginning of our observations and ends at the end of the observation period, Fig. (\ref{fig:main_scheme}). 

How long is the longest time that we can reasonably consider? We take the longest known inter-glitch period, $\approx 284\,\ds$, and imagine that for scenario 1) r-mode emission starts with our observations and lasts $(284-50)\,\ds$, at which time the next glitch happens; for scenario 2) r-mode emission starts $(284-50-\tauCW)\,\ds$ before the beginning of the observation period (where $\tauCW$ is the duration of the gravitational-wave observation in days). In these two cases the inter-glitch epochs, i.e. the mid-times in-between two successive glitches, are 
\vspace{-.5em}
\begin{equation}
    \label{eq:def_midepoch}
	\begin{cases}
     	t_1 = t^\textrm{gw}_0 + 92 ~\textrm{days}\\
   	t_2 = t^\textrm{gw}_0 + \tauCW - 142 ~\textrm{days}, \\ 
	\end{cases}
\end{equation}
with $t^\textrm{gw}_0$ being the time corresponding to the start of the gravitational-wave observation period and the subscripts ``1" and ``2"  indicating the two different scenarios. Consistently with \citep{Antonopoulou:2017hwa} we can then determine the signal parameters at $t_1$ and $t_2$ by evolving the values of Tab.~\ref{tab:RotEvolution} defined at the last observed inter-glitch epoch $t_\textrm{obs}$ with $\nLT, \ddnuLT$. We obtain $\nu|^\textrm{LT}_{t_1}$ and $\nu|^\textrm{LT}_{t_2}$. Since $\nu|^\textrm{ST}_{t_1} \equiv \nu|^\textrm{LT}_{t_1}, \, \nu|^\textrm{ST}_{t_2} \equiv \nu|^\textrm{LT}_{t_2}$, 
we use these values to derive the ones at the reference time for each search $\Tref=t^\textrm{gw}_0+ {1\over2}\,\tauCW$, by evolving them according to the short term (ST) evolution model of Tab.~\ref{tab:RotEvolution}. Specifically, we consider $\nig$, $\ddnuig^\min $ and $\ddnuig^\max$ and from these, using the definition of braking index, we derive $\dot\nu(t_1,\nig,\ddnuig^\min)$, $\dot\nu(t_2,\nig,\ddnuig^\min)$, $\dot\nu(t_1,\nig,\ddnuig^\max)$ and $\dot\nu(t_2,\nig,\ddnuig^\max)$. We evolve these to $\Tref$ and find four values of $\nu(\Tref)$ and four values of $\dot\nu(\Tref)$, corresponding to $(t_1,\ddnuig^\min)$, $(t_2,\ddnuig^\min)$, $(t_1,\ddnuig^\max)$ and $(t_2,\ddnuig^\max)$. We take $\nu_\min$, $\nu_\max$, $\dot\nu_\min$ and $\dot\nu_\max$ as the smallest and largest among the four. These quantities define  
the range of possible spin frequencies and derivatives. We assume that $\ddot\nu_\min=\ddnuig^\min$ and $\ddot\nu_\max=\ddnuig^\max$.  

\begin{deluxetable}{llccc}[tbh!]
\tablecaption{The range of spin frequency and frequency-derivatives for $\PSR$ at the reference time of each search. The parameter uncertainties from Table \ref{tab:RotEvolution} are propagated throughout the derivations described in the text and are indicated with brackets. \label{tab:NumValNu}}
\tablehead{ 
	\multicolumn{2}{c}{\TBstrut Search run} &  \colhead{O1}   & \colhead{\TBstrut  O2.1\TBstrut }  &  \colhead{\TBstrut  O2.2 \TBstrut }
} 
\startdata 
	\TBstrut $\nu_\min$ & [$\Hz$] &  \numinOOne & \numinOTwoOne & \numinOTwoTwo  \\  
	\TBstrut $\nu_\max$ & [$\Hz$] & \numaxOOne & \numaxOTwoOne & \numaxOTwoTwo  \\ \hline 
	\TBstrut $\dnu_\min$ & [$\HzS$] & \multicolumn{3}{c}{ \nudotmin }  \\ 
	\TBstrut $\dnu_\max$ & [$\HzS$] & \multicolumn{3}{c}{ \nudotmax } \\ \hline 
	\TBstrut $\ddnu_\min$ & [$\HzSsq$] & \multicolumn{3}{c}{ \nuddotmin }  \\ 
	\TBstrut $\ddnu_\max$ & [$\HzSsq$] & \multicolumn{3}{c}{ \nuddotmax } 
\enddata  
\end{deluxetable}

The reference times for each observation period O1, O2.1 and O2.2 are given in Table~\ref{tab:GridSpacings} and the corresponding boundaries for the spin frequency and spindown in Table~\ref{tab:NumValNu}: 

All these values can be substituted in Eq.~(\ref{eq:searchRanges}) and finally yield the gravitational-wave frequency and frequency derivative search ranges shown in Table~\ref{tab:SearchParams}. 

\newpage
\begin{deluxetable}{llccc}[tbh!]
\tablecaption{The search parameter space. The brackets indicate uncertainties.\label{tab:SearchParams}}
\tablehead{ 
	\multicolumn{2}{c}{\TBstrut Search run} &  \colhead{O1}   & \colhead{\TBstrut  O2.1\TBstrut }  &  \colhead{\TBstrut  O2.2 \TBstrut }
} 
\startdata 
\TBstrut $\fmin$ & [$\Hz$] &  $\fminOOne$ & $\fminOTwoOne$ & $\fminOTwoTwo$ \\ 
\TBstrut $\fmax$ & [$\Hz$] &  $\fmaxOOne$ & $\fmaxOTwoOne$ & $\fmaxOTwoTwo$ \\ \hline
\TBstrut $\fBand$ &  [$\Hz$]  &  $\fbandOOne$ &  $\fbandOTwoOne$ &  $\fbandOTwoTwo$ \\ \hline
\TBstrut $\dfmin$ & [$\HzS$]  & \multicolumn{3}{c}{ $\fdotmin$ } \\ 
\TBstrut $\dfmax$ & [$\HzS$]  & \multicolumn{3}{c}{ $\fdotmax$ } \\ \hline
\TBstrut $\dfBand$ & [$\HzS$]  & \multicolumn{3}{c}{ $\fdotband$ } \\ \hline
\TBstrut $\ddfmin$ & [$\HzSsq$]  & \multicolumn{3}{c}{ $\fddotmin$ } \\ 
\TBstrut $\ddfmax$ & [$\HzSsq$]  & \multicolumn{3}{c}{ $\fddotmax$ } \\ \hline
\TBstrut $\ddfBand$ & [$\HzSsq$]  & \multicolumn{3}{c}{ $\fddotband$ } \\ \hline
\enddata  
\end{deluxetable}

\subsection{Detection statistics} 
\label{subsec:DetMethod}

We perform a fully coherent, multi-detector search using a maximum likelihood matched filtering method known as $\Fstat$ \citep{Cutler:2005hc}. The $\Fstat$ is the optimal frequentist statistic for this type of signal, in the presence of stationary, Gaussian detector noise. The resulting detection values, $2\F$, for each template represent the likelihood that a signal with the template's waveform be present in the data, with respect to Gaussian noise. In Gaussian noise $2\F$ follows a $\chisq$-distribution with $4$ degrees of freedom ($\chisqdegr$) and a non-centrality parameter that equals the squared signal-to-noise ratio \citep{Jaranowski:1998qm}. 

In the presence of spectral disturbances in the data, the detection statistic can be improved by extending the noise model to include ``noise lines'', on top of Gaussian noise \citep{Keitel:2013wga}. The corresponding line-robust statistic $\BSGL$ requires the choice of the tuning parameter $\scFtho$ that defines the single-detector $\Fstat$ magnitude at which $\BSGL$ begins to down-rank search results with respect to the pure Gaussian noise model. $\scFtho$ is usually defined in terms of a Gaussian-noise false-alarm probability $\chi^2(2\scFtho | 0)$, which we choose to be $\approx 1/\Neff$, the effective  number of independent templates. The results of the search are ranked according to $\BSGL$. For this search we estimate that $\Neff = 0.9 \, N_\textrm{tot}$, where $N_\textrm{tot}$ is the total number of searched templates.

\subsection{The search set up}
\label{subsec:SearchSetup}

The search targets different wave shapes, each defined by a value of the gravitational-wave frequency and frequency-derivatives, $f,\fdot,\fddot$. The ensemble of waveforms obtained by varying the values for the $f,\fdot,\fddot$ within the boundaries given by Table~\ref{tab:SearchParams} , constitutes the signal template bank of our search. 

\begin{deluxetable}{llccc}[tbh!]
\tablecaption{Parameters of each search, including the template grid spacings, the start and reference times $\Tst$ and $\Tref$, the search time-baseline $\Tcoh$, the total time for which there is data from both detectors $\Tdata$ (expressed as the total number of input SFTs), an estimate of the number of independent templates $\Neff$ and the tuning parameter $\scFtho$. \label{tab:GridSpacings}}  
\tablehead{ 
	\multicolumn{2}{c}{\TBstrut Search run} & \colhead{O1} & \colhead{O2.1 }  &  \colhead{O2.2}
} 
\startdata 
	\TBstrut $\Tst$  & [GPS] & $\TstOOne$ & $\TstOTwoOne$ & $\TstOTwoTwo$ \\ \hline
	\TBstrut $\Tref$ & [GPS]  & $\TrefOOne$ & $\TrefOTwoOne$ & $\TrefOTwoTwo$ \\ \hline
	\TBstrut $\Tcoh$ & [$\ds$] & $\tauOOne$ & $\tauOTwoOne$ & $\tauOTwoTwo$ \\ \hline
	\TBstrut $\Tdata$ & [$\Nsft$] & $\TdataOOne$ & $\TdataOTwoOne$ & $\TdataOTwoTwo$ \\ \hline
	\TBstrut $\dF$ & [$\Hz$] & $\dFOOne$ & $\dFOTwoOne$ & $\dFOTwoTwo$ \\ \hline
	\TBstrut $\ddF$ & [$\HzS$]  & $\ddFOOne$ & $\ddFOTwoOne$ & $\ddFOTwoTwo$ \\ \hline
	\TBstrut $\dddF$ & [$\HzSsq$] & $\dddFOOne$ & $\dddFOTwoOne$ & $\dddFOTwoTwo$ \\ \hline
	\multicolumn{2}{l}{\TBstrut $\log_{10} \Neff$} & $\lgNeffOOne$ & $\lgNeffOTwoOne$ & $\lgNeffOTwoTwo$ \\ \hline
	\multicolumn{2}{l}{\TBstrut $\scFtho$} & $\FstarOOne$ & $\FstarOTwoOne$ & $\FstarOTwoTwo$ \\
\enddata  
\end{deluxetable}  

\begin{figure}[tbh!]
    \centering
    \includegraphics[width=.9\columnwidth]{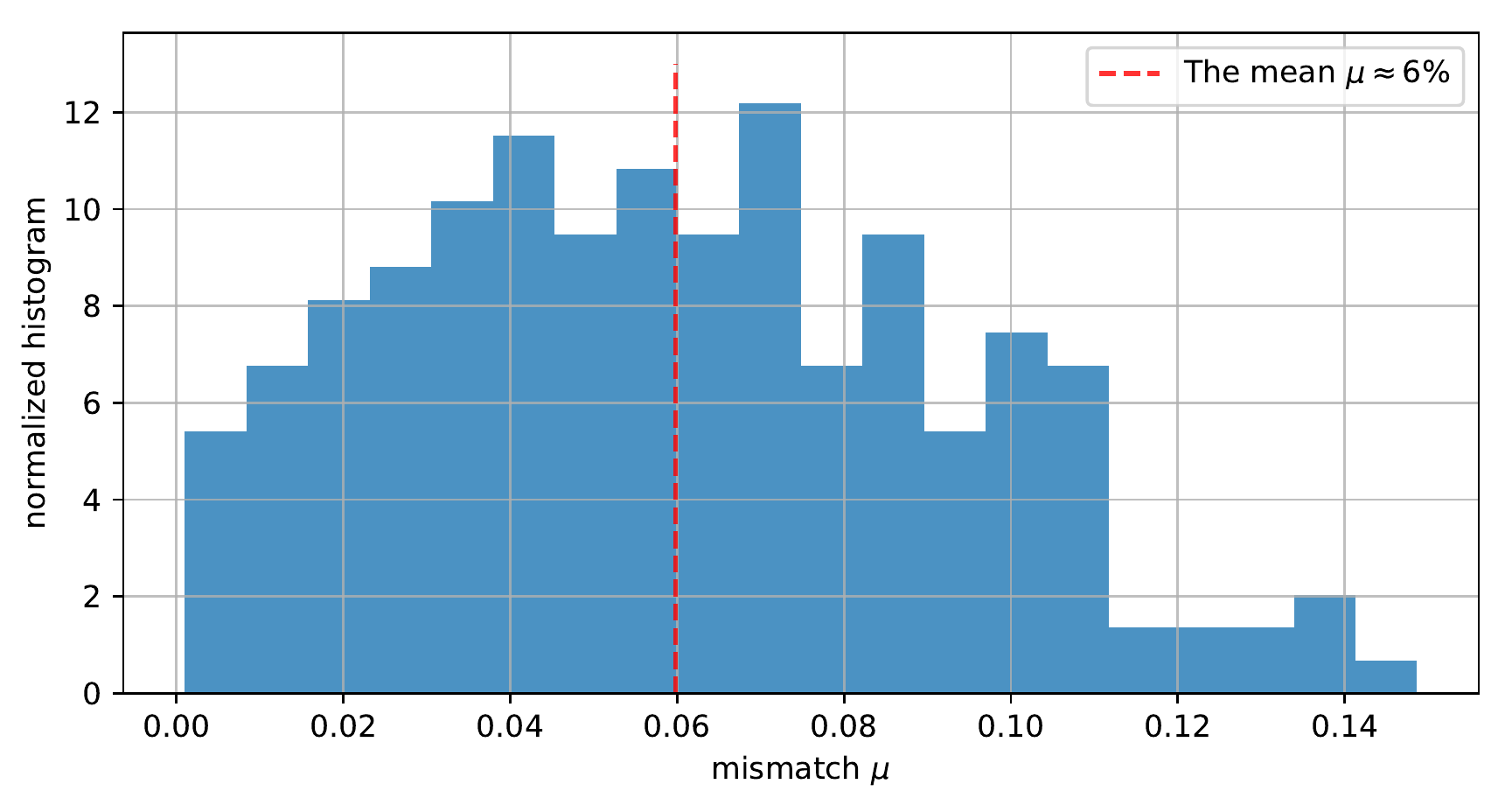}
    \caption{The mismatch distribution $\mu$ for the search template grid computed from $200$ searches on fake signals. $\mu={{\rho^2_\textrm{PM}-\rho^2_\textrm{grid}}\over{\rho^2_\textrm{PM}}}$ with $\rho^2$ being the signal-to-noise measured with a perfect match between signal and template (``PM") and with a search over the original search grid (``grid").}
    \label{fig:mismatch}
\end{figure}

The grid spacings $(\dF, \delta\dot f, \dddF)$ are such that the average loss in detection efficiency due to signal-template mismatch is about $6\%$. The mismatch distribution is shown Fig.~\ref{fig:mismatch}. The details of the procedure can for instance be found in \citep{Behnke:2014tma}. Since the $\delta\ddot f$ spacing is smaller than $\ddot f_\min$ from Eq.s~\ref{eq:searchRanges}, we set $\ddot f_\min =0$.

A summary of all search parameters is given in Table ~\ref{tab:GridSpacings}. Overall, we search $\approx 10^{13}$ templates in every search.

\section{Results}
\label{sec:results}

\begin{figure}[tbh!]
    \centering
    \includegraphics[width=\columnwidth]{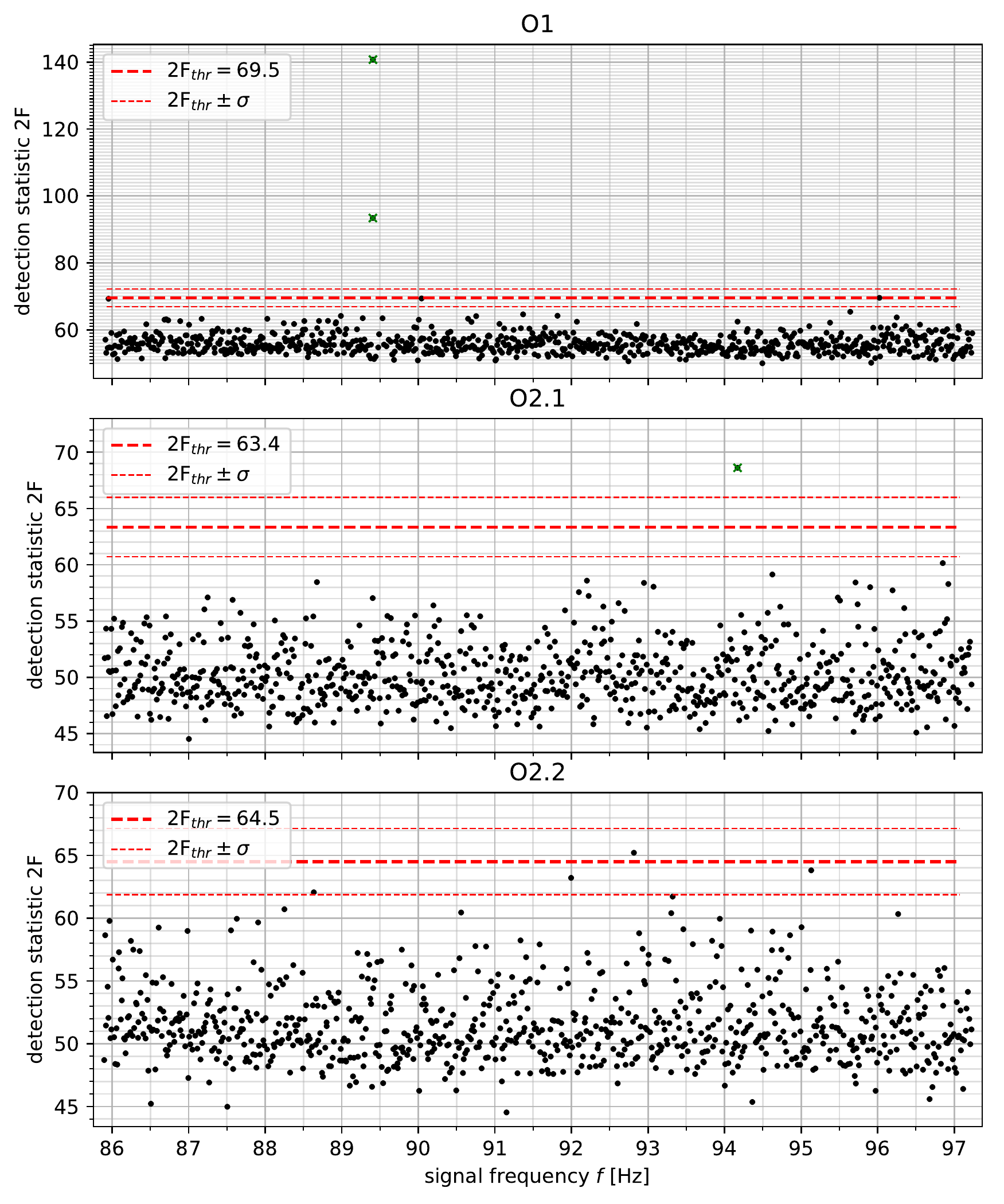}
    \caption{The most significant candidate in every $0.014$ Hz band. We note the value of $2\F_\textrm{thr}$ for each search, the expected highest detection statistic value in Gaussian noise over the number of independent templates searched.}
    \label{fig:search_output}
\end{figure}

Fig.~\ref{fig:search_output} shows the $2\F$ values of the most significant $\BSGL$ results in every 0.014 Hz band, after having excluded results close to spectral artefacts that were cleaned-out in the input data \citep{Covas:2018oik,Abbott:2017pqa}. The cleaning procedure substitutes the real data with fake Gaussian noise, hence these data do not contribute any astrophysical information to the results. The extent of the excluded region is determined by the extent of the original spectral contamination and by an evaluation of the spread that this would generate in signal-frequency space, for our specific target and for the largest searched $|\dot f|$. This removes $3.6\%$ (O1), $2.2\%$ (O2.1) and $3\%$ (O2.2) of the results. The complete list of the bands excluded from further inspection and from the upper limit statements, is given in appendix \ref{app:excludedBands}.

We compare the $2\F$ values with $2\F_\textrm{thr}$, the expected most significant $2\F$ over the entire search in Gaussian noise:
\begin{equation}
	2\F_\textrm{thr}=\int_0^\infty \chi^2_4 ~\Neff ~ F_{\chi^2_4}^{(\Neff-1)} ~ p_{\chi^2_4}~d\chi^2_4,
	\label{eq:pdf_loudest}
\end{equation}
where $ p_{\chi^2_4}$ is the probability density function of a $\chi^2_4$ variable and $F_{\chi^2_4}$ its cumulative distribution. We do not use $\Fthr$ as a rigorous measure of significance but rather as an indicator.

If a candidate were found well above $\Fthr$ with consistent parameters across the three searches, this would {\it not} automatically mean that it is a signal from \Jpsr, but it would certainly warrant further investigation. On the other hand, if no consistent candidate exists above the expected loudest, it is unlikely that we can confidently identify a signal with this search.

\begin{deluxetable}{cccccc}  
\tablecaption{\label{tab:Candidates} The most significant candidates from each of the searches.}  
\tablehead{ 
   \colhead{$f$ [$\Hz$]} & \colhead{$\dot{f}$ [$\HzS$]} & \colhead{$\ddot{f}$ [$\HzSsq$]} & \colhead{$2\F$} & \colhead{$\frac{2\F-\Fthr}{\sigma}$} & \colhead{$\BSGL$} 
} 
\startdata 
\multicolumn{6}{c}{\TBstrut \textbf{O1}} \\ \hline  
\TBstrut 89.4071 & $-2.72 \times 10^{-10}$ & $1.96 \times 10^{-20}$ & 140.7 & 27.0 & 0.77 \\ 
\TBstrut 96.0229 & $-6.34 \times 10^{-10}$ & $9.70 \times 10^{-21}$ & 69.5 & 0.0 & 0.52 \\ 
\TBstrut 85.9534 & $-3.41 \times 10^{-10}$ & $9.70 \times 10^{-21}$ & 69.3 & -0.1 & 0.46 \\ 
\TBstrut 90.0425 & $-4.57 \times 10^{-10}$ & $2.95 \times 10^{-20}$ & 69.4 & -0.1 & 0.44 \\ 
\TBstrut 89.4060 & $-3.40 \times 10^{-10}$ & $9.70 \times 10^{-21}$ & 93.4 & 9.0 & -0.02 \\ 
\hline \multicolumn{6}{c}{\TBstrut \textbf{O2.1}} \\ \hline  
\TBstrut 94.1700 & $-4.72 \times 10^{-10}$ & $1.96 \times 10^{-20}$ & 68.6 & 2.0 & 1.67 \\ 
\hline \multicolumn{6}{c}{\TBstrut \textbf{O2.2}} \\ \hline  
\TBstrut 92.8159 & $-5.00 \times 10^{-10}$ & $1.96 \times 10^{-20}$ & 65.2 & 0.3 & 0.69 \\ 
\TBstrut 95.1321 & $-6.83 \times 10^{-10}$ & $1.96 \times 10^{-20}$ & 63.8 & -0.3 & 0.38 \\ 
\TBstrut 91.9957 & $-6.19 \times 10^{-10}$ & $1.96 \times 10^{-20}$ & 63.2 & -0.5 & 0.25 \\ 
\TBstrut 88.6347 & $-4.71 \times 10^{-10}$ & $1.96 \times 10^{-20}$ & 62.1 & -0.9 & 0.00 \\ 
\enddata  
\end{deluxetable}

We find ten candidates with $2\F\geq 2\F_\textrm{thr}-\sigma$, and they are listed in Table~\ref{tab:Candidates}. We comment only on the three that exceed at least the $+1\sigma$ level.

\begin{figure*}[tbh!]
\centering 
\subfloat[Candidate in O1 at $f=89.4071\,\Hz$]{%
         \includegraphics[width=.4\textwidth]{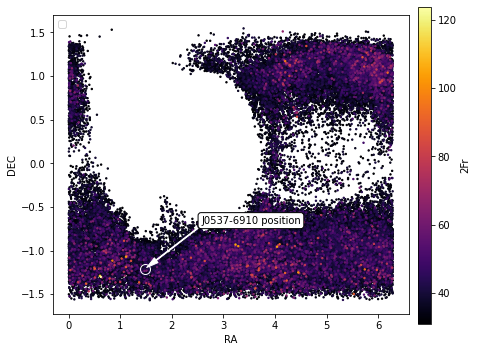}
         \includegraphics[width=.4\textwidth]{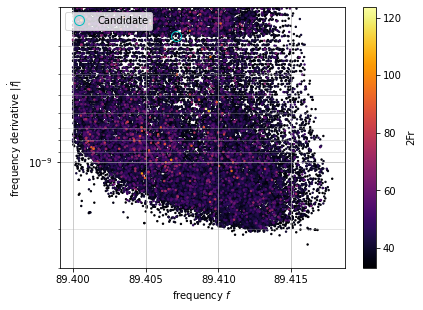}
  \label{fig:ASO1_cand1}
}\qquad
\subfloat[Candidate in O2.1 at $f=94.17\,\Hz$]{%
         \includegraphics[width=.4\textwidth]{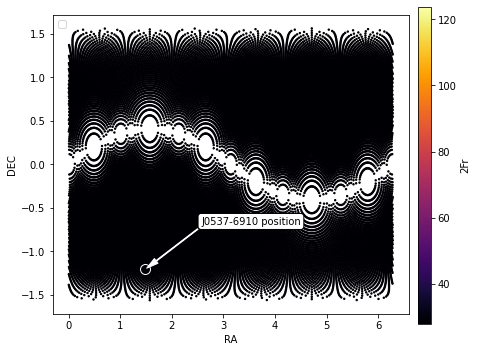}
         \includegraphics[width=.4\textwidth]{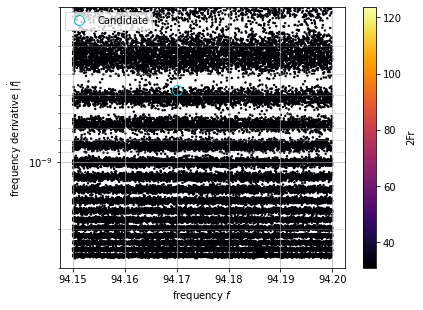}
  \label{fig:ASO2_1_cand1}
}
\caption{Results of the all-sky O1 and O2 searches \citep{Abbott:2017pqa,O2ASEaH} in the parameter regions of the two candidates of Table \ref{tab:Candidates}. The plots show the detection statistic values (color-coded) as a function of the template-waveform frequency and frequency derivative in the top panels, and as a function of the template-waveform source position $(\alpha, \delta)$ in the bottom panels. When a spectral region is contaminated, the distribution of candidates is not uniform in parameter space, and this can be clearly seen in panel (a), as opposed to panel (b) that portrays results from an undisturbed frequency region.}
\end{figure*}

{\bf{O1 search}}: There are two outstanding candidates at $\approx 89.41\,\Hz$ whose detection  statistic exceeds the expectation for the loudest by about 9$\sigma$ and 27$\sigma$, respectively. Their proximity in frequency indicates that the two candidates are due to the same root cause, which we identify in spectral disturbances in the Hanford detector. This is also clearly reflected in the results of an all-sky search on the same data \citep{Abbott:2017pqa}: the distribution over sky position of the top results from that search is typical of a disturbance rather than of a signal or of Gaussian noise, and the results at the position of \Jpsr\ are not significant in any way. Figure~\ref{fig:ASO1_cand1} illustrates our findings. 

{{\bf{O2.1 search}}: The most significant candidate at $94.17\,\Hz$ is $\approx 2\sigma$ above the expectation. The all-sky search around this frequency doesn't reveal any disturbance, as shown in Figure~\ref{fig:ASO2_1_cand1}. The average noise of the detectors at the relevant frequencies does not exhibit any notable feature.

{\bf{O2.2 search}}: There are no significant candidates from this search: all the listed candidates are within $1\sigma$ of the expectations.

\subsection{Upper limits on the gravitational-wave amplitude} 
\label{subsec:ULs_h0}

\begin{figure}[tbh!]
    \centering
    \includegraphics[width=\columnwidth]{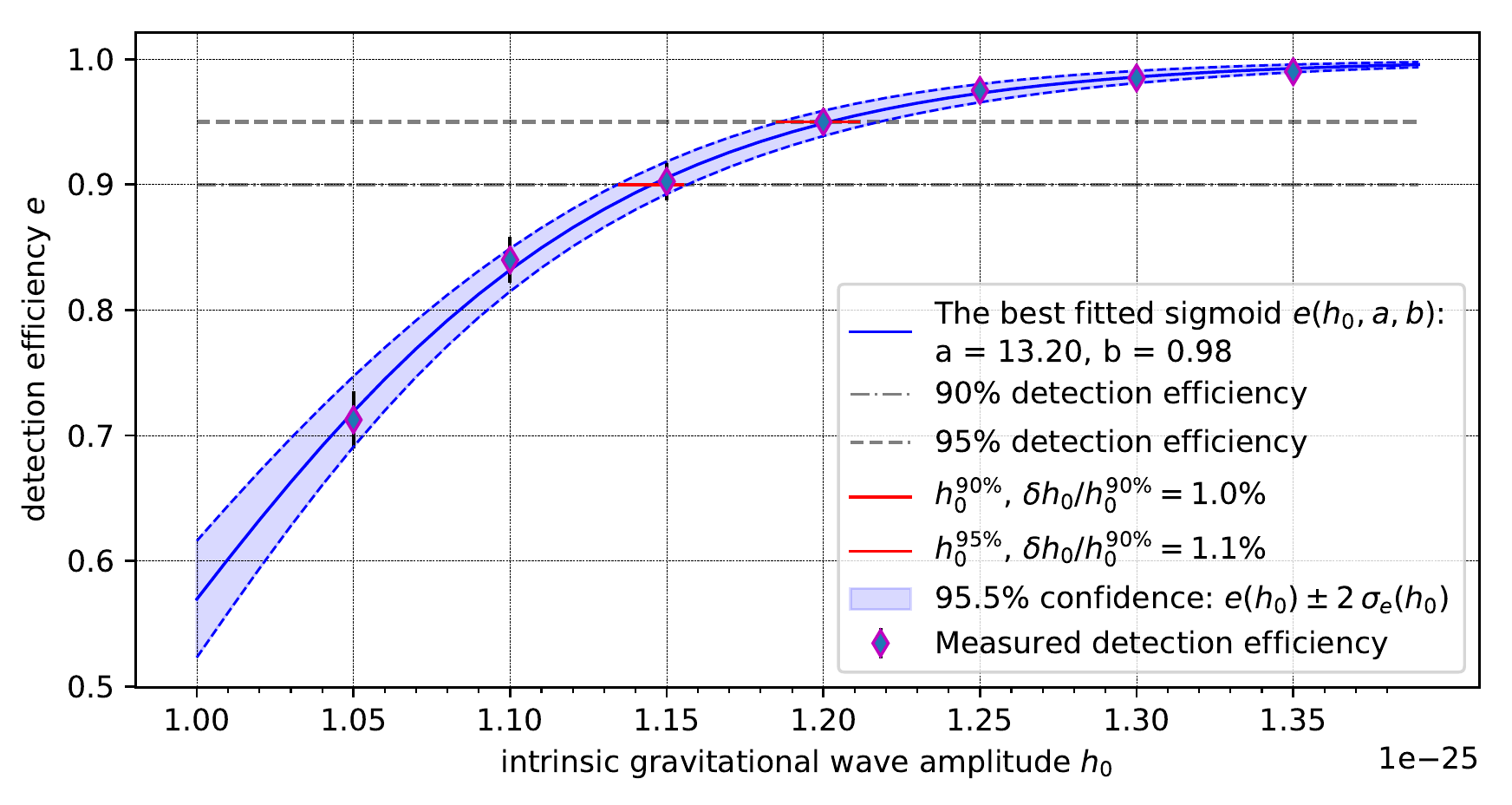}
    \caption{Detection efficiency curve for the band $87.75-88.25$ Hz in O2.1 search run. This is a typical result.}
    \label{fig:95ConfidenceRegionExample}
\end{figure}

\begin{figure}[tbh!]
    \centering
    \includegraphics[width=\columnwidth]{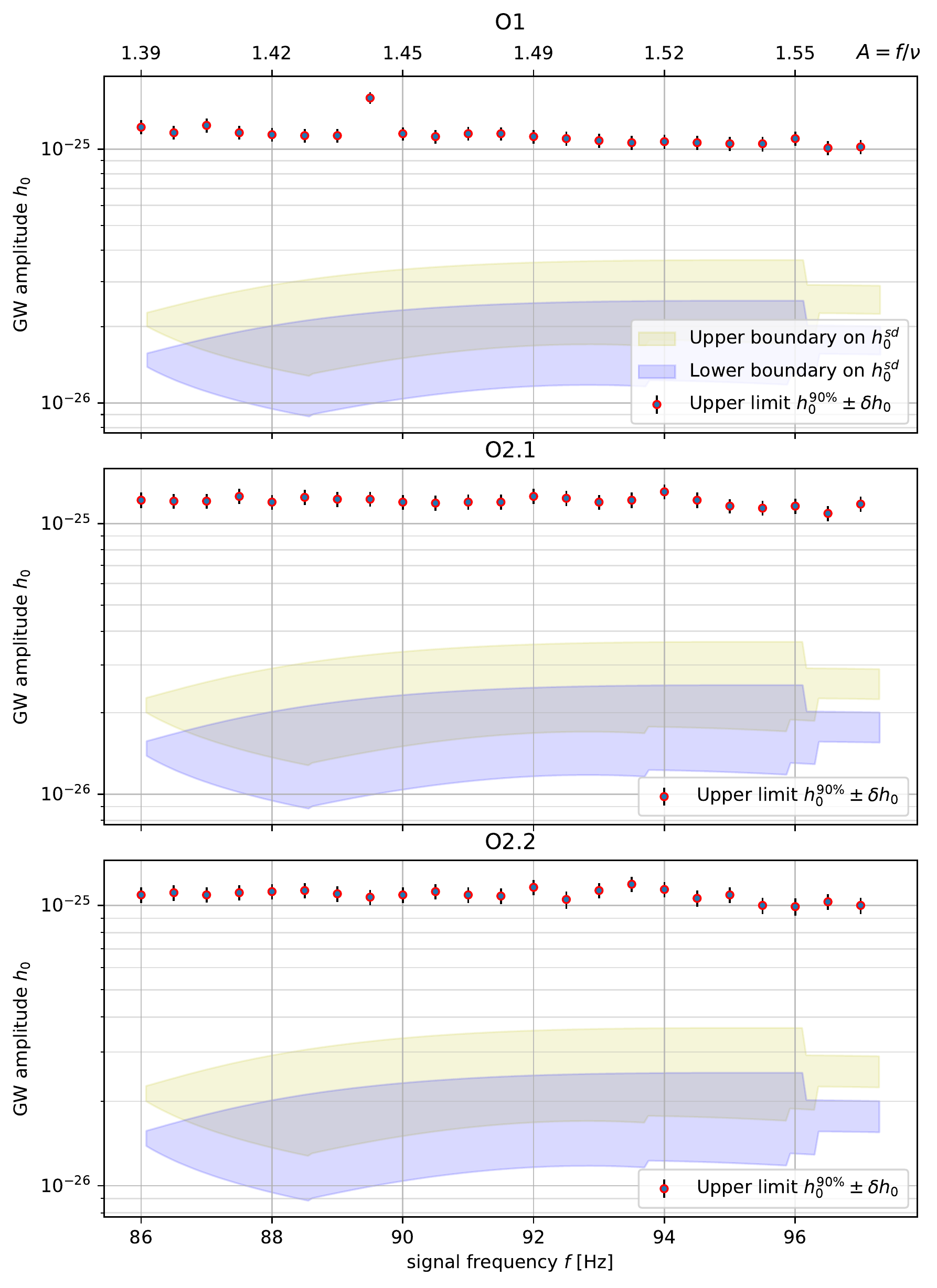}
    \caption{The markers show the upper limits on gravitational-wave amplitude $\hUL$ for a continuous signal from \Jpsrs from each of the searches.The shaded regions show the range of values that the spin-down upper limit could take, depending on the equation of state of the the star, as described in Section \ref{subsec:sdlimit}. The second x-axis, on the top, shows $A(f)$ for the ${\hsd}_{\max}$ curve, i.e. $A=f/\nu_\min$. On this scale it is however not possible to appreciate the difference with  $A=f/\nu_\max$, so the plotted axes hold for all quantities shown.}
    \label{fig:h0ULs}
\end{figure}

We set upper limits on the intrinsic gravitational amplitude $h_0$ in 0.5 Hz bands, based on the highest detection statistic value measured in each band, after the results from the fake Gaussian noise bands are removed. We perform 200 fake-signal search-and-recovery Monte Carlos within each band. The signals are all at the location of \Jpsr, with frequency, spindown and initial phase values taken from uniform random distributions in their respective ranges. We add these signals in the real data. We consider values of $h_0$ ranging from $9.5\ee{-26}$ to $1.8\ee{-25}$. 

The searches are performed with the same grids and set-up as the search, Tab. \ref{tab:SearchParams}, in the neighbourhood of the fake signal parameters. A signal is counted as recovered if the highest detection statistic value from the fake-signal search is higher than the one recorded in the actual search. The detection efficiency $e(h_0)$ is the fraction of recovered signals.

We adopt a sigmoid of the form $e(h_0)=(1+\exp({{{\textrm{a}}-h_0}\over{\textrm{b}}}))^{-1}$ to fit $h_0$ with the corresponding measured detection efficiency. We use Python's ``curve fit'' package \citep{curvefit} based on the Levenberg-Marquardt algorithm through the least squares method. 
The uncertainties in $e$ stemming from the measurement error on the number of recovered signals are translated in  
uncertainties on the fit parameters $\delta a$ and  $\delta b$, computed as the square root of the diagonal elements of the covariance matrix. We use $\delta a$ and  $\delta b$ to estimate the standard deviation $\sigma_e(h_0)$ of the best fit sigmoid $e(h_0)$. Fig.~\ref{fig:95ConfidenceRegionExample} shows an example of the sigmoid fit with two curves $e(h_0)\pm 2 \, \sigma_e(h_0)$ that bracket the expected $e(h_0)$ curve with $> 95\%$ confidence. 

The 90\% confidence upper limit on the intrinsic gravitational-wave amplitude is the smallest $h_0$ such that $90\%$ of the target signal population in the search range would have produced a value of the detection statistic higher than the one that was measured in the search. We read this value, $\hUL$, off the sigmoid fit curve at $e=0.9$. 

The uncertainty $\delta e$  determines the range of variability for $\hUL$ which overall amounts to $\le 2\%$ of the upper limit value. We add the calibration uncertainty which we conservatively take to be $5\%$ \citep{Cahillane:2017vkb}. The upper limits together with their uncertainties are plotted in Fig.~\ref{fig:h0ULs} for all 3 search runs. They are provided in tabular form in the appendix \ref{app:ULsData} and in machine-readable format at \citep{machineReadables}. We also compute the 95\% confidence upper limits, which are $\lesssim$ 5\% higher than the $90\%$ confidence ones. 

\subsubsection{Sensitivity depth}
\label{sec:sensDepth}

\begin{figure}[tbh!] 
   \includegraphics[width=\columnwidth]{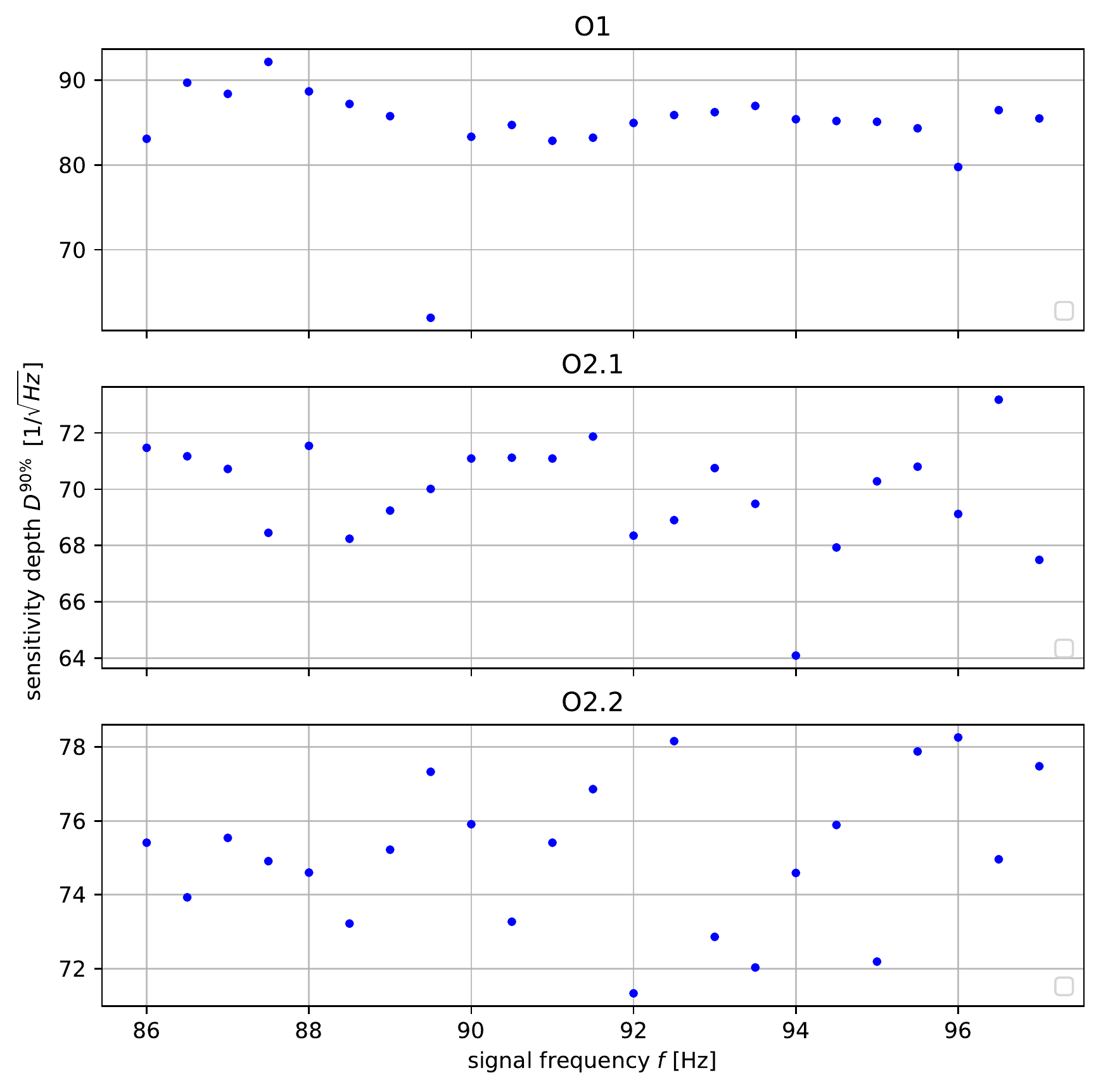}
\caption{The sensitivity depth}  
\label{fig:SensDepth}
\end{figure}

The sensitivity depth is a useful measure to compare the baseline performance of different searches \citep{Dreissigacker:2018afk}. It was first introduced in \citep{Behnke:2014tma} as
\begin{equation}
{{\mathcal{D}}}^{90\%}(f)={\sqrt{S_h(f)}\over {\hUL(f) }}~~[ {1/\sqrt{\text{Hz}}} ],
\label{eq:sensDepth}
\end{equation}
where $\sqrt{S_h(f)}$ is the noise level associated with the signal frequency $f$.  
The multi-detector $S_h(f)$ for our searches is the harmonic mean of the single-detector power spectral densities $S_h^H$ and $S_h^L$ of the data, then averaged over the 0.5 Hz frequency band that the upper limit value refers to. The resulting ${{\mathcal{D}}}^{90\%}(f)$ is shown in Fig.~\ref{fig:SensDepth} and tabulated in Appendix~\ref{app:ULsData}. We provide the values of $\sqrt{S_h(f)}$ in machine readable format at  \citep{machineReadables}.

\subsubsection{Spin-down limit}
\label{subsec:sdlimit}

If all the kinetic energy lost by \Jpsrs (its spin-down) rotating at $\nu$ is due to gravitational emission at frequency $f$, its gravitational-wave amplitude is

\begin{equation}
	\hsd = {1\over D}\sqrt{{{10 G}\over c^3} \Izz {{|\dnu| \nu}\over f^2}~} 
\label{eq:h0General_sd1} 
\end{equation}
where $\Izz$ is the moment of inertia of the star with spin axis in the $\hat z$ direction. If in Eq.s~\ref{eq:CWfreq_rmode} we neglect the terms in $(\nu/\nuK)^2$ (slowly rotating star) and set $A = f/\nu$ then  
\vspace{-.2em}
\begin{equation}
	\hsd = {1\over {A\:D}}\sqrt{{{10 G}\over c^3} \Izz \dnuovnu} .
\label{eq:h0General_sd2} 
\end{equation}
This is a general formula that applies to any emission mechanism. If the emission is due to an equatorial ellipticity in the star, then $A=2$ and we find the commonly-seen spindown-limit formula, for example Eq.~5 of \citep{Aasi:2013sia}. 

In the case of r-mode emission $A$ encodes information on the equation of state of the star. As shown in Fig.~\ref{fig:UniversalFit}, mass $M$ and radius $R$ are different functions of $A$ for different equations of state. If $\mathcal{C}=M/R$ is the compactness of the star, then
$A = | -1.373 + .079\,\mathcal{C} - 2.25\,\mathcal{C}^2| \in [1.39; 1.57]$ for $M \in [1.02 - 2.76\,\Msun]$ and compactness $\mathcal{C}=M/R \in [0.11, 0.31]$. This was found by fitting $14$ realistic equations of state by \citep{Idrisy:2014qca} and we will use it in Eq.~\ref{eq:h0spindown} to compute $M(A)$ and $R(A)$ from $M(\mathcal{C})$ and $R(\mathcal{C})$ given in \citep{Ozel:2015fia, NSEoS}.

The moment of inertia $\Izz$ also depends on the equation of state. We re-write it in terms of the normalized moment of inertia $\Bar{I} := \Izz/M^3$ that can be expressed in terms of $\mathcal{C}$ for slowly rotating stars with the coefficients given in Tab. 2 of \citep{Breu:2016ufb}. 

Eq. (\ref{eq:h0General_sd2}) then becomes
\vspace{-.2em}
\begin{equation}
	\hsd(A, \dnuovnu) = {1\over {A\:D}}\sqrt{{{10 G}\over c^3} \Bar{I}(A) {M^3(A)} \dnuovnu}.
\label{eq:h0spindown}
\end{equation}
We consider two extremes: 
\begin{equation}
	\label{eq:h0sdRanges}
	\begin{cases}
    	{\hsd}_{\min}(A)=\hsd(A, {{|\dot\nu|}\over \nu}|_\min),~~{{|\dot\nu|}\over \nu}|_\min=|\dnu|_\min/\nu_\max \\ 
    	{\hsd}_{\max}(A)=\hsd(A,  {{|\dot\nu|}\over \nu}|_\max),~{{|\dot\nu|}\over \nu}|_\max=|\dnu|_\max/\nu_\min. 
	\end{cases}
\end{equation}
As $M(A)$ varies in the range shown in middle panel of Fig.~\ref{fig:UniversalFit}, we find the corresponding range $\Delta{\hsd}_{\min}$ and $\Delta{\hsd}_{\max}$. We set  $A=f/\nu_\min$ in $\Delta{\hsd}_{\max}$ and  $A=f/\nu_\max$ in $\Delta{\hsd}_{\min}$ and derive the two differently shaded regions of Fig.~\ref{fig:h0ULs} which define range of variability of the spin-down upper limit $\hsd$. We are neglecting the $B \nu^2/\nuK^2$ term of Eq.~\ref{eq:CWfreq_rmode} for simplicity. This approximation is completely unimportant in the context of sketching the boundaries of $\hsd$. 
\begin{figure}[tbh!]
    \centering
    \includegraphics[width=\columnwidth]{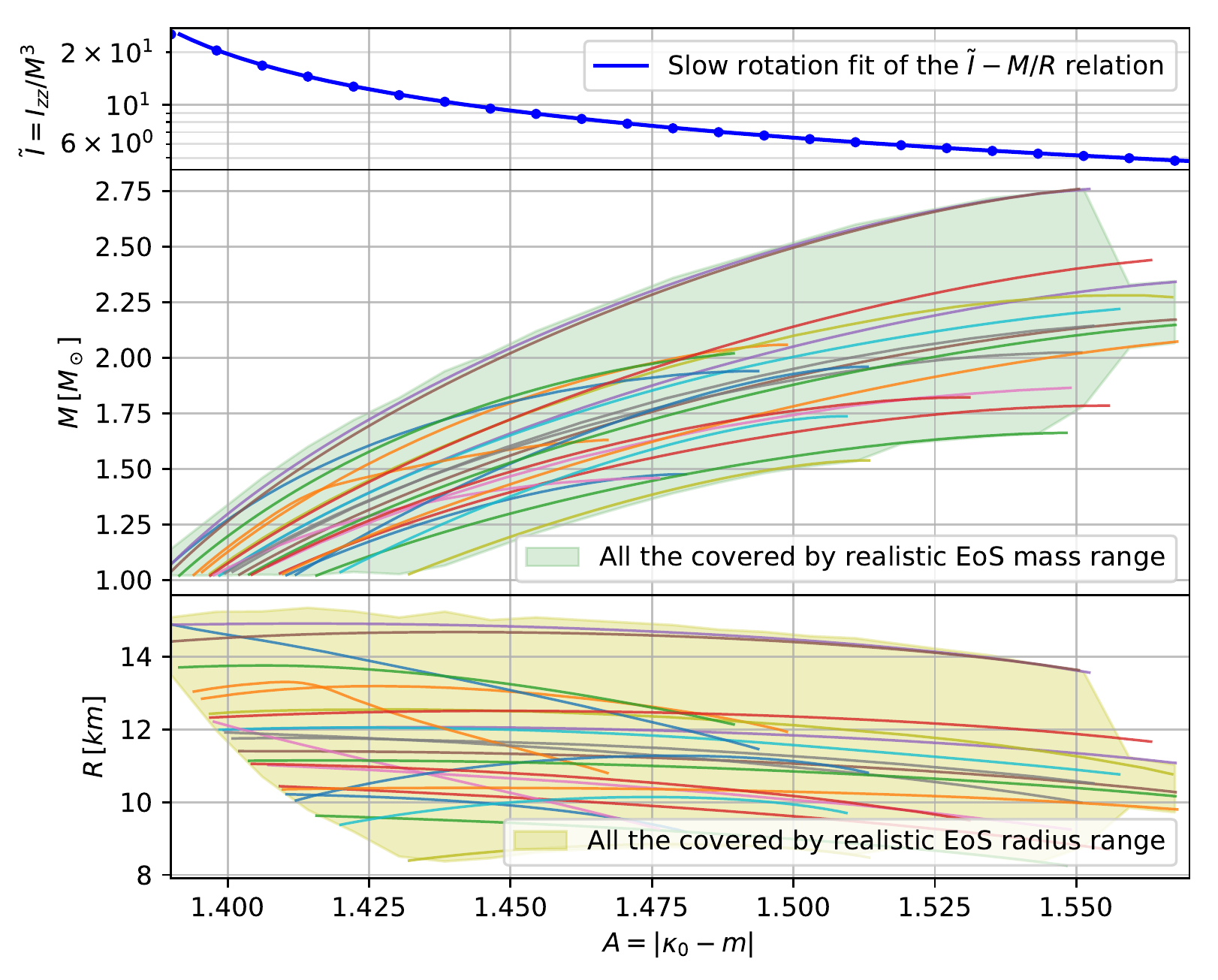}
    \caption{$\Tilde{I}-M/R$ Universal relation and ranges of $M$ and $R$ from realistic EoS.}
    \label{fig:UniversalFit}
\end{figure}

\subsection{Upper limits on the r-mode amplitude}

The r-mode saturation amplitude $\alpha$ that supports gravitational-wave emission with a strain $h_0$ at a frequency $f$ from a source at a distance $D$ is \citep{Owen:2010ng}:
\begin{equation}
    \alpha = \sqrt{\frac{5}{8\pi}} {c^5\over G} \frac{h_0}{(2\pi f)^3} \frac{D}{MR^3 \Tilde{J}},
    \label{eq:rmodes}
\end{equation}
where $\Tilde{J}$ is the dimensionless canonical angular momentum of the r-mode \citep{Owen:1998xg}. $\Tilde{J}$ is less dependant on the equation of state than $M$ and $R$ so, following  \citep{Owen:1998xg}, we fix its value to $0.0164$ (as computed from a polytropic EoS with index $n=1$) and encapsulate the dependancy on the equation of state in the term $MR^3$, as function of $A$.

The gravitational-wave frequency also depends on $A$: $f=A\nu$. As done is Section \ref{subsec:sdlimit} we consider
\vspace{-.2em}
\begin{equation}
	\label{eq:alphaRanges}
	\begin{cases}
    	{\alpha}^{\min}(h_0,A,f=A\nu_\max)= \sqrt{\frac{5}{8\pi}} {c^5\over G} \frac{h_0}{(2\pi A\nu_\max)^3} \frac{D}{M(A)R(A)^3 \Tilde{J}}\\ 
    	{\alpha}^{\max}(h_0,A,f=A\nu_\min)=\sqrt{\frac{5}{8\pi}} {c^5\over G} \frac{h_0}{(2\pi A\nu_\min)^3} \frac{D}{M(A)R(A)^3 \Tilde{J}}
	\end{cases}
\end{equation}
and for each of these curves the range $A \in [1.39,\,1.57]$ determines the range of variability of the saturation amplitude $\alpha$ as a function of $h_0$. 
In practice since ${\alpha}^{\min}(h_0,A,f=A\nu_\max)\approx {\alpha}^{\max}(h_0,A,f=A\nu_\min)$ we convert the gravitational-wave amplitude upper-limits $\hUL$ to ranges for the r-mode amplitude upper limits in every half $\Hz$ bands as 
\begin{equation}
	\label{eq:alphaULs}
	\alpha^{90\%}(f) = {\alpha}(\hUL,A,f=A\nu_\min),~ A \in [1.39,\,1.57]. 
	\end{equation}
The results are shown on Fig. (\ref{fig:alphaULs}) for all search runs. The shaded area represent the spread of $\alpha$ in the possible range of $M(A) R(A)^3$ bounded by realistic equations of state, as well as the upper limit for $M=1.4\,\Msun,\, R=11.7$ km (middle black curve).

\begin{figure}[h!tbp]
   \includegraphics[width=\columnwidth]{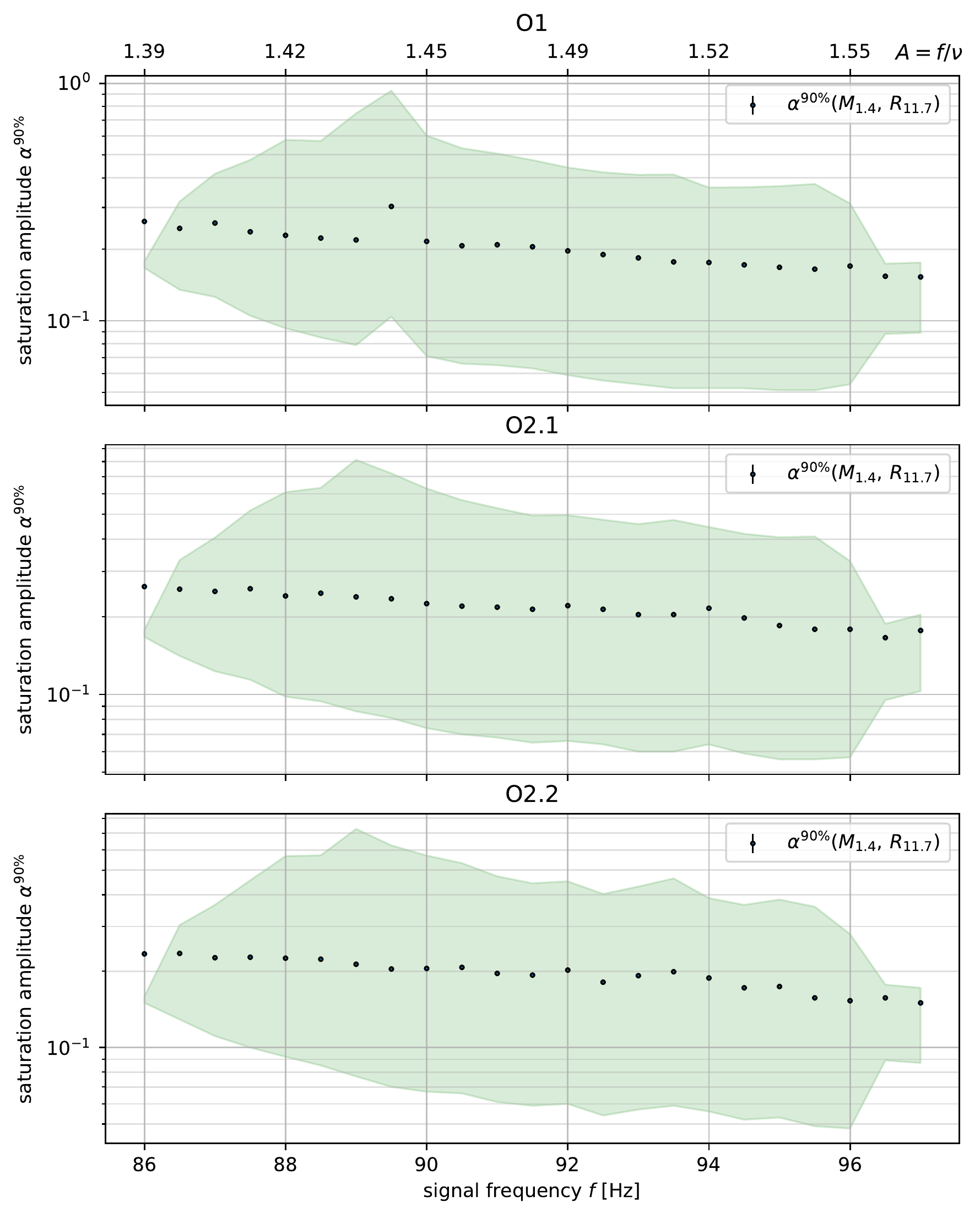}
    \caption{Upper limits on the r-mode saturation amplitude $\alpha$ derived from the gravitational-wave amplitude upper limits $\hUL$.}  
    \label{fig:alphaULs}
\end{figure}

\subsection{Not always ``ON" signal}
\label{subsec:transientSignal}

Our upper limits are based on the optimistic assumption that the r-mode signal is always ``ON" during the  time of the searches. This might not be the case because in the model that we consider, r-mode emission begins some time after a glitch and ends with the next glitch. Not knowing when glitches happened for \Jpsrs around the O1 and O2 observing times, we cannot be sure that some of our search times do not fall in a period too close to a glitch to be emitting r-modes, according to our model. In order to estimate the impact of this assumption we randomly pick start times for the O1 and O2 runs during the 13 years for which we have glitch-occurrence times and based on this glitch-time, we compute the fraction of these simulated  O1, O2.1 and O2.2 runs which overlaps with the r-mode emission period\footnote{We recall that we have defined the r-mode emission period to be the period 50-day after a glitch to the next glitch.\label{note1}}. The resulting distributions for 1000 draws of the start times are shown in the top panel of Fig.~\ref{fig:ULsFraction}. Since the data has gaps which are not uniformly distributed, the fraction of the overlapping {\it{time}} is not equal to the fraction of {\it{data}} in the overlap stretches, so we also compute this and show the distributions in the bottom panel of  Fig.~\ref{fig:ULsFraction}. We find that the 50th percentiles for the overlap fraction of the data are $\approx 45\%$ for O1 and $\approx 50\%$ for  O2.1 and O2.2. We repeat the simulation-and-search Monte Carlos described in Section \ref{subsec:ULs_h0} for O1, O2.2 and O2.2 with signals from this population and with frequency between $87.75$ Hz and $88.25$ Hz. In this sample frequency band we find a $\hUL$ higher by a factor $\approx 4.4, 3.7$ and $4.2$ respectively for the three searches, compared to the always-ON-signal results. 

\section{Conclusions}
\label{sec:conclusions}

\begin{figure*}[tbh!]
\centering 
\subfloat[]{%
	\includegraphics[width=.69\textwidth]{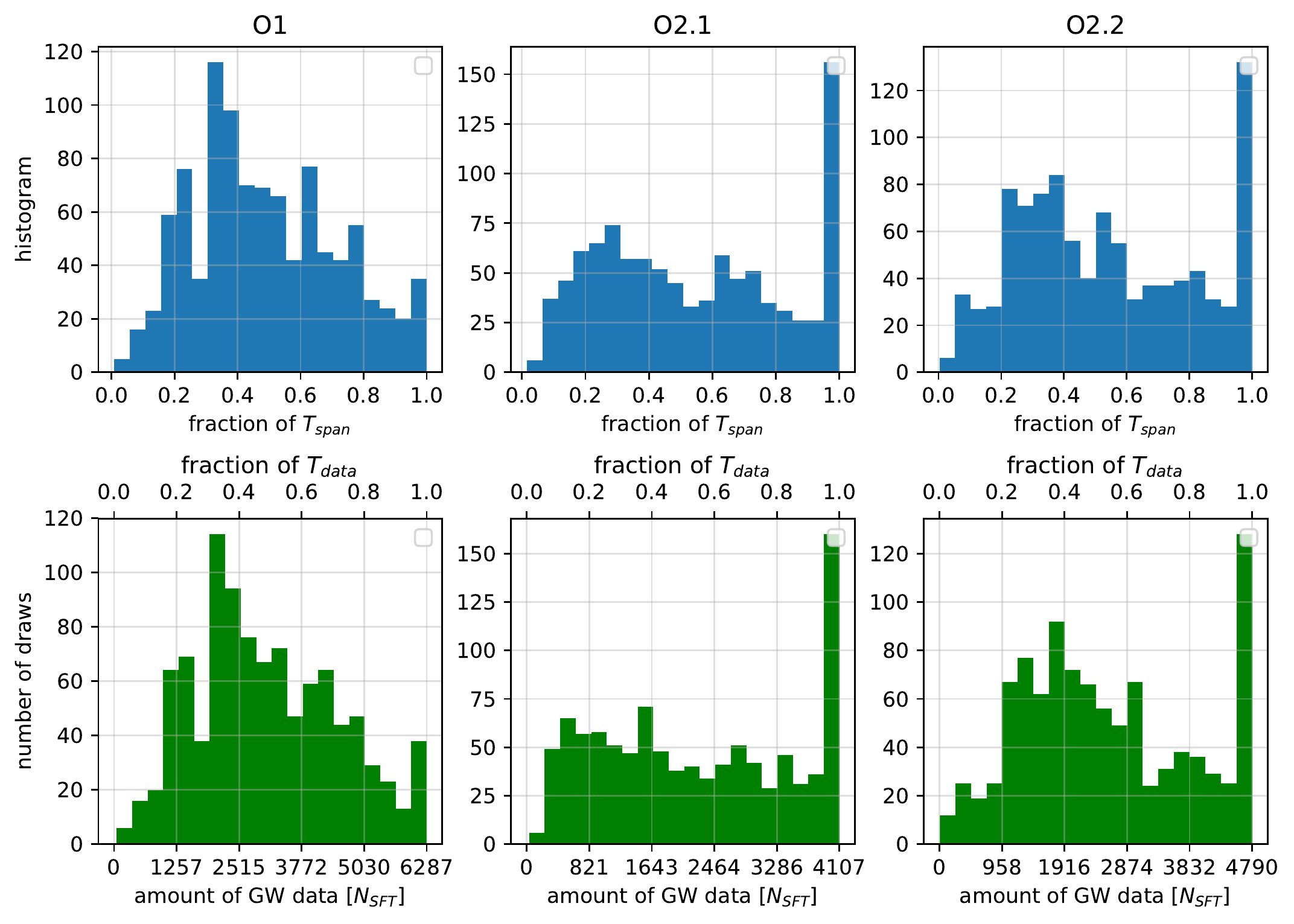}
	\label{fig:Fraction_CWsearch}
}\qquad
\subfloat[]{%
	\includegraphics[width=.26\textwidth]{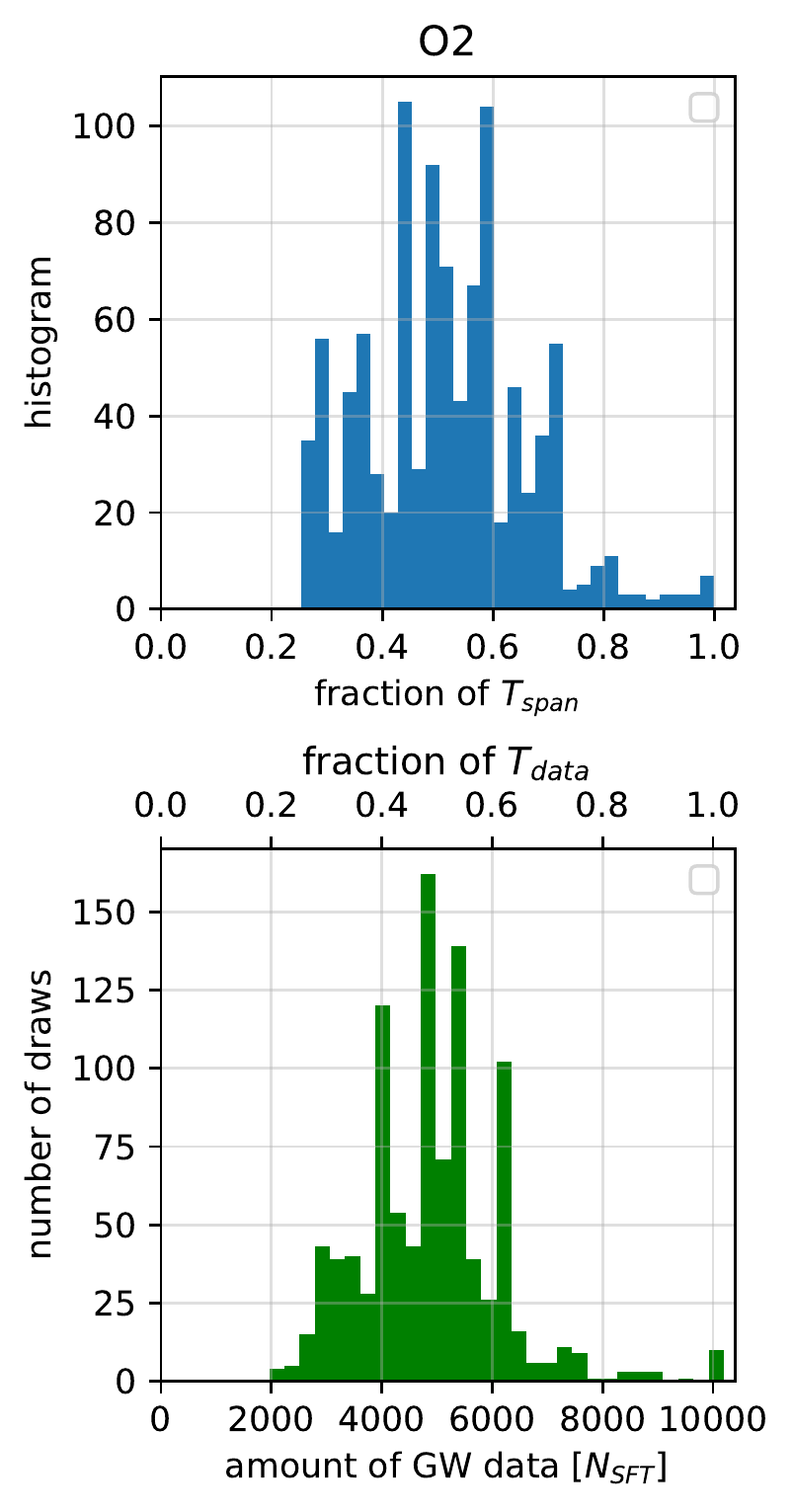}
	\label{fig:Fraction_O2}
}
\caption{Distributions of the fraction of the searches' observation spans $\tauCW$ (top panels) and distribution of the amount of gravitational-wave data $T_\textrm{data}$ (lower panels) that overlap with unstable r-mode emission periods (for O1, O2.1 and O2.2) and that overlaps with the longest glitch-free period (for O2). 
        The amount of data is expressed by the number of input SFTs ($\Nsft$). The total number of SFTs is given in Tab. \ref{tab:GridSpacings} as $\Tdata$ for O1, O2.1, O2.2, and  $\Tdata = 10194\,\Nsft$ for O2.
        These r-mode emission periods are simulated by drawing 1000 random search start-times during the 13 years for which we have the occurrence-time of \Jpsrs glitches.}
        \label{fig:ULsFraction}
\end{figure*}

Pulsar \Jpsrs is an intriguiging candidate for r-mode gravitational-wave emission $\gtrsim$ 50 days after a glitch. Unfortunately we do not know whether the object glitched during the O1 and O2 LIGO data runs, so we carry out three coherent searches for r-mode continuous gravitational waves on periods lasting several tens of days. We choose the periods based on the available data and its gaps. We pick the frequency and frequency-derivative range  to be large enough to include all uncertainties in the rotation frequency timing model, evolved to the time of the observations, and the uncertainties stemming from the unknown equation of state of the star. 
This is the first search specifically targeting r-mode emission from a known pulsar. While we do not find evidence for a continuous gravitational-wave signal, we report a marginal outlier from the O2.1 search at $\sim 94.17$ Hz. We set upper limits on the gravitational-wave amplitude of r-mode signals in 0.5 Hz bands. Overall, the $h_0$ upper limits span a range between $(1-1.6)\times 10^{-25}$, with an average sensitivity depth between $75$ and $85\, {1\over{\sqrt{\Hz}}}$, consistent with the size of the data sets employed. The r-mode saturation amplitude values that this search could detect are consistent with those necessary to interpret the EM observations in terms of unstable r-mode emission \citep{Andersson:2017fow}. They are about an order of magnitude larger than the physically most plausible ones but scenarios can be imagined where even such high values are possible. We refer the reader to the discussion on this point in Section 3 of \citep{Andersson:2017fow}.  Our upper limits are a factor of $\approx$ 5 higher than the average spin-down limit amplitude.

Lacking precise ephemeris data for this pulsar, \Jpsrs was not included in the LIGO 220+ known-pulsars search \citep{Authors:2019ztc}, but was later targeted in a small, 0.25 Hz search around twice the rotation frequency, 123.86 Hz, in \citep{Abbott:2019bed}. That search, carried out on the O2 data, is limited to a spin-down range of $8 \times10^{-13}$ Hz/s and overall comprises $\sim 1.6\times 10^{9}$ templates, about 10000 times fewer than employed here for each of our searches. The upper limits of \citep{Abbott:2019bed} are consistent with the longer coherent time-baseline, the different level of the detector noise and the significantly smaller template bank of that search with respect to the one presented here. 

As in \citep{Abbott:2019bed}, our upper limits are based on the optimistic assumption that the r-mode signal is always ``ON" during the  time of the searches. Based on historical glitch-occurrence times we construct a population of signals with varying durations and overlaps with our data-sets, and evaluate the upper-limits on this population of signals. The sensitivity is degraded with respect to the always-ON population by a factor $\approx 4$.

The likelihood of a glitch occurring during the 232 days observation time  of the LIGO search \citep{Abbott:2019bed} is reflected in a mean overlap of the observation time with the longest inter-glitch periods of $50\%$ (see Fig.~\ref{fig:ULsFraction}). \citep{Abbott:2019bed} do not comment on their loss of sensitivity due to a possible glitch of \Jpsr. We estimate that for a realistically glitching signal the upper limit $\hUUL$ of \citep{Abbott:2019bed} would be $\approx 3.6$ times higher, comparable to the degradation that we report for our searches. Timing of \Jpsrs in order to identify the times when glitches occur, eliminates all these uncertainties and is of paramount importance to search for continuous gravitational waves from \Jpsr. 

A very interesting candidate from this search would be a high-significance signal consistent in at least two of the three searches. This would indicate a repeating phenomenon exciting the star's r-mode instability, coherent with the observations of  \citep{Andersson:2017fow}. A definitive confirmation would need a verification on a different gravitational-wave data set, corroborated by glitch information from EM observations. This re-inforces the importance of EM timing of \Jpsr.

A detection would be of great importance for multiple reasons. It would be i) the first detection of a continuous gravitational-wave signal, opening interesting prospects for high-precision tests of gravity ii) the first direct observation of gravitational-waves emission through unstable r-modes, as predicted in \citep{Andersson:1997xt}  iii) the discovery that at least some young neutron stars loose angular momentum due to r-mode gravitational waves iv) a probe of neutron star interior.

As new and more sensitive gravitational-wave data becomes available, deeper searches will be possible, also including the use of specific techniques on longer data sets \citep{Keitel:2015ova,Keitel:2018pxz,Ashton:2018qth}. The scientific return of gravitational-wave searches like this is greatly enhanced when timing data is available, that identifies the rotation parameters during the gravitational-wave observations and glitch-occurrence times. NICER \citep{nicer} could provide this invaluable information to the broad scientific community.

\section{Acknowledgments}
We thank Wynn Ho for useful discussions in the early phases of this project, Danai Antonopoulou for having supplied us with the numerical values of Table 1 and 2 of \citep{Antonopoulou:2017hwa}, Benjamin Steltner for having prepared the input data for this search, and Heinz-Bernd Eggenstein and the Einstein@Home team for access to the all-sky search results. We acknowledge David Keitel for his comments on our presentation at the 30th Texas Symposium in Portsmouth. A special thank-you to Nils Andersson for the many interesting discussions and for his comments on the manuscript. 
\par\noindent
This research has made use of data and web tools for data transfer of the gravitational-wave Open Science Center \citep{LOSC}, a service of LIGO Laboratory, the LIGO Scientific Collaboration and the Virgo Collaboration. LIGO is funded by the U.S. National Science Foundation. Virgo is funded by the French Centre National de Recherche Scientifique (CNRS), the Italian Istituto Nazionale della Fisica Nucleare (INFN) and the Dutch Nikhef, with contributions by Polish and Hungarian institutes.

\bibliography{refs}{}
\bibliographystyle{aasjournal}

\appendix
\section{Excluded frequency bands}
\label{app:excludedBands}

The tables in this appendix contain the complete list of frequency bands excluded from the analysis and from the upper limit results.

\setcounter{table}{0}
\renewcommand{\thetable}{\Alph{table}1} 
\begin{table*}[tbh!]
\centering
   \caption{Excluded frequency bands in the O1 search run}
   \label{tab:LinesO1}
  \begin{minipage}[t]{.3\textwidth} 
\begin{tabular}{ccc} 
\hline \multicolumn{3}{c}{\TBstrut Hanford (LHO)} \\ \hline 
central $f$  &  $- \Delta f$ & $+ \Delta f$  \\ 
${\scriptstyle [\Hz]}$ & ${\scriptstyle [\Hz]}$ & ${\scriptstyle [\Hz]}$ \\ \hline 
$86.000000$ & $0.004856$ & $0.004360$ \\ 
$86.500000$ & $0.004856$ & $0.004360$ \\ 
$87.000000$ & $0.004856$ & $0.004360$ \\ 
$87.500000$ & $0.004856$ & $0.004360$ \\ 
$88.000000$ & $0.004356$ & $0.003860$ \\ 
$88.000000$ & $0.004856$ & $0.004360$ \\ 
$88.500000$ & $0.004856$ & $0.004360$ \\ 
$89.000000$ & $0.004856$ & $0.004360$ \\ 
$89.500000$ & $0.004856$ & $0.004360$ \\ 
$90.000000$ & $0.004856$ & $0.004360$ \\ 
$90.500000$ & $0.004856$ & $0.004360$ \\ 
$91.000000$ & $0.004856$ & $0.004360$ \\ 
$91.500000$ & $0.004856$ & $0.004360$ \\ 
$92.000000$ & $0.004856$ & $0.004360$ \\ 
$92.500000$ & $0.004856$ & $0.004360$ \\ 
$93.000000$ & $0.004856$ & $0.004360$ \\ 
$93.500000$ & $0.004856$ & $0.004360$ \\ 
$94.000000$ & $0.004856$ & $0.004360$ \\ 
$94.238100$ & $0.006856$ & $0.006360$ \\ 
$94.244700$ & $0.006856$ & $0.006360$ \\ 
$94.500000$ & $0.004856$ & $0.004360$ \\ 
$95.000000$ & $0.004856$ & $0.004360$ \\ 
$95.500000$ & $0.004856$ & $0.004360$ \\ 
$96.000000$ & $0.004356$ & $0.003860$ \\ 
$96.000000$ & $0.004856$ & $0.004360$ \\ 
$96.500000$ & $0.004856$ & $0.004360$ \\ 
$97.000000$ & $0.004856$ & $0.004360$ \\ 
\hline \end{tabular} 
\end{minipage} 
\begin{minipage}[t]{.3\textwidth} 
\begin{tabular}{ccc} 
\hline \multicolumn{3}{c}{\TBstrut Livingston (LLO)} \\ \hline 
central $f$  &  $- \Delta f$ & $+ \Delta f$  \\ 
${\scriptstyle [\Hz]}$ & ${\scriptstyle [\Hz]}$ & ${\scriptstyle [\Hz]}$ \\ \hline 
$86.749750$ & $0.004861$ & $0.004371$ \\ 
$87.749725$ & $0.004861$ & $0.004371$ \\ 
$87.900000$ & $0.004361$ & $0.003871$ \\ 
$88.400000$ & $0.004361$ & $0.003871$ \\ 
$88.749700$ & $0.004861$ & $0.004371$ \\ 
$89.749675$ & $0.004861$ & $0.004371$ \\ 
$90.300000$ & $0.004361$ & $0.003871$ \\ 
$90.749650$ & $0.004861$ & $0.004371$ \\ 
$90.800000$ & $0.004361$ & $0.003871$ \\ 
$91.300000$ & $0.004361$ & $0.003871$ \\ 
$91.749625$ & $0.004861$ & $0.004371$ \\ 
$92.749600$ & $0.004861$ & $0.004371$ \\ 
$93.700000$ & $0.004361$ & $0.003871$ \\ 
$93.749575$ & $0.004861$ & $0.004371$ \\ 
$94.200000$ & $0.004361$ & $0.003871$ \\ 
$94.749550$ & $0.004861$ & $0.004371$ \\ 
$95.749525$ & $0.004861$ & $0.004371$ \\ 
$95.883160$ & $0.010261$ & $0.006571$ \\ 
$96.600000$ & $0.004361$ & $0.003871$ \\ 
$96.749500$ & $0.004861$ & $0.004371$ \\ 
$97.100000$ & $0.004361$ & $0.003871$ \\ 
\hline \end{tabular} 
\end{minipage} 

\end{table*}
\vspace{3em}

\setcounter{table}{0}
\renewcommand{\thetable}{\Alph{table}2} 
\vspace{5em}
\begin{table*}[tbh!]
   \caption{Excluded frequency bands in the O2.1 and O2.2 search runs}
   \label{tab:LinesO3}
   \begin{minipage}[t]{.45\textwidth} 
\begin{tabular}{ccccc} 
\hline \multicolumn{5}{c}{\TBstrut Hanford (LHO)} \\ \hline 
central $f$ & $- \Delta f$ & $+ \Delta f$ & $- \Delta f$ & $+ \Delta f$ \\ 
${\scriptstyle [\Hz]}$ & ${\scriptstyle [\Hz]}$ & ${\scriptstyle [\Hz]}$ & ${\scriptstyle [\Hz]}$ & ${\scriptstyle [\Hz]}$ \\ \hline 
$85.998700$ & $0.001781$ & $0.001974$ & $0.002815$ & $0.002971$ \\ 
$86.000000$ & $0.001781$ & $0.001974$ & $0.002815$ & $0.002971$ \\ 
$86.500000$ & $0.003641$ & $0.003834$ & $0.004675$ & $0.004831$ \\ 
$86.749837$ & $0.001781$ & $0.001974$ & $0.002815$ & $0.002971$ \\ 
$86.998700$ & $0.001781$ & $0.001974$ & $0.002815$ & $0.002971$ \\ 
$87.000000$ & $0.001781$ & $0.001974$ & $0.002815$ & $0.002971$ \\ 
$87.500000$ & $0.003641$ & $0.003834$ & $0.004675$ & $0.004831$ \\ 
$87.749822$ & $0.001781$ & $0.001974$ & $0.002815$ & $0.002971$ \\ 
$87.998700$ & $0.001781$ & $0.001974$ & $0.002815$ & $0.002971$ \\ 
$88.000000$ & $0.001781$ & $0.001974$ & $0.002815$ & $0.002971$ \\ 
$88.500000$ & $0.003641$ & $0.003834$ & $0.004675$ & $0.004831$ \\ 
$88.749806$ & $0.001781$ & $0.001974$ & $0.002815$ & $0.002971$ \\ 
$88.889400$ & $0.001781$ & $0.001974$ & $0.002815$ & $0.002971$ \\ 
$88.889840$ & $0.001781$ & $0.001974$ & $0.002815$ & $0.002971$ \\ 
$88.998700$ & $0.001781$ & $0.001974$ & $0.002815$ & $0.002971$ \\ 
$89.000000$ & $0.001781$ & $0.001974$ & $0.002815$ & $0.002971$ \\ 
$89.500000$ & $0.003641$ & $0.003834$ & $0.004675$ & $0.004831$ \\ 
$89.749791$ & $0.001781$ & $0.001974$ & $0.002815$ & $0.002971$ \\ 
$89.998700$ & $0.001781$ & $0.001974$ & $0.002815$ & $0.002971$ \\ 
$90.000000$ & $0.001781$ & $0.001974$ & $0.002815$ & $0.002971$ \\ 
$90.500000$ & $0.003641$ & $0.003834$ & $0.004675$ & $0.004831$ \\ 
$90.749775$ & $0.001781$ & $0.001974$ & $0.002815$ & $0.002971$ \\ 
$90.998700$ & $0.001781$ & $0.001974$ & $0.002815$ & $0.002971$ \\ 
$91.000000$ & $0.001781$ & $0.001974$ & $0.002815$ & $0.002971$ \\ 
$91.160252$ & $0.005621$ & $0.005814$ & $0.006655$ & $0.006811$ \\ 
$91.500000$ & $0.003641$ & $0.003834$ & $0.004675$ & $0.004831$ \\ 
$91.749760$ & $0.001781$ & $0.001974$ & $0.002815$ & $0.002971$ \\ 
\hline \end{tabular} 
\end{minipage} 
\begin{minipage}[t]{.45\textwidth} 
\begin{tabular}{ccccc} 
\hline \multicolumn{5}{c}{\TBstrut Livingston (LLO)} \\ \hline 
central $f$ & $- \Delta f$ & $+ \Delta f$ & $- \Delta f$ & $+ \Delta f$ \\ 
${\scriptstyle [\Hz]}$ & ${\scriptstyle [\Hz]}$ & ${\scriptstyle [\Hz]}$ & ${\scriptstyle [\Hz]}$ & ${\scriptstyle [\Hz]}$ \\ \hline 
$91.998700$ & $0.001781$ & $0.001974$ & $0.002815$ & $0.002971$ \\ 
$92.000000$ & $0.001781$ & $0.001974$ & $0.002815$ & $0.002971$ \\ 
$92.500000$ & $0.003641$ & $0.003834$ & $0.004675$ & $0.004831$ \\ 
$92.749745$ & $0.001781$ & $0.001974$ & $0.002815$ & $0.002971$ \\ 
$92.998700$ & $0.001781$ & $0.001974$ & $0.002815$ & $0.002971$ \\ 
$93.000000$ & $0.001781$ & $0.001974$ & $0.002815$ & $0.002971$ \\ 
$93.500000$ & $0.003641$ & $0.003834$ & $0.004675$ & $0.004831$ \\ 
$93.749729$ & $0.001781$ & $0.001974$ & $0.002815$ & $0.002971$ \\ 
$93.998700$ & $0.001781$ & $0.001974$ & $0.002815$ & $0.002971$ \\ 
$94.000000$ & $0.001781$ & $0.001974$ & $0.002815$ & $0.002971$ \\ 
$94.500000$ & $0.003641$ & $0.003834$ & $0.004675$ & $0.004831$ \\ 
$94.749714$ & $0.001781$ & $0.001974$ & $0.002815$ & $0.002971$ \\ 
$94.998700$ & $0.001781$ & $0.001974$ & $0.002815$ & $0.002971$ \\ 
$95.000000$ & $0.001781$ & $0.001974$ & $0.002815$ & $0.002971$ \\ 
$95.500000$ & $0.003641$ & $0.003834$ & $0.004675$ & $0.004831$ \\ 
$95.749698$ & $0.001781$ & $0.001974$ & $0.002815$ & $0.002971$ \\ 
$95.998700$ & $0.001781$ & $0.001974$ & $0.002815$ & $0.002971$ \\ 
$96.000000$ & $0.001781$ & $0.001974$ & $0.002815$ & $0.002971$ \\ 
$96.500000$ & $0.003641$ & $0.003834$ & $0.004675$ & $0.004831$ \\ 
$96.749683$ & $0.001781$ & $0.001974$ & $0.002815$ & $0.002971$ \\ 
$96.998700$ & $0.001781$ & $0.001974$ & $0.002815$ & $0.002971$ \\ 
\hline \end{tabular} 
\end{minipage} 
 
\end{table*}
\vspace{5em}

\section{Upper limits on the gravitational wave and the saturation amplitude}
\label{app:ULsData}

Table \ref{tab:UppLim} shows that there are upper limits in every 0.5 Hz band. The central frequency f of each band is indicated in the first column. We stress again that the $\hUL$ and $\alphaUL$ upper limits do not hold for the subbands of Appendix \ref{app:excludedBands}.

\setcounter{table}{1}
\renewcommand{\thetable}{\Alph{table}1} 
\begin{deluxetable*}{ccc|cccc|cccc|cccc} 
\tablecaption{Upper Limits on the Gravitational-wave and Saturation Amplitudes \label{tab:UppLim}} 
\tablehead{ 
	 \multicolumn{3}{c|}{\TBstrut Spin-down limit} & \multicolumn{4}{c|}{O1} & \multicolumn{4}{c|}{O2.1} & \multicolumn{4}{c}{O2.2} \\ \hline  
	 \colhead{$f$} & \colhead{${\hsd}_\min$} & \colhead{${\hsd}_\max$} & \colhead{$\hUL$} & \colhead{${\alpha}^{90\%}$} & \colhead{${\sqrt{S_h}}$} & \colhead{${\mathcal{D}}^{90\%}$} & \colhead{$\hUL$} & \colhead{${\alpha}^{90\%}$} & \colhead{${\sqrt{S_h}}$} & \colhead{${\mathcal{D}}^{90\%}$} & \colhead{$\hUL$} & \colhead{${\alpha}^{90\%}$} & \colhead{${\sqrt{S_h}}$} & \colhead{${\mathcal{D}}^{90\%}$} \\  
	 \colhead{${\scriptstyle [\Hz]}$} & \colhead{${\scriptstyle \left [10^{-26} \right ]}$} & \colhead{${\scriptstyle \left [10^{-26} \right ]}$} & \colhead{${\scriptstyle \left [10^{-25} \right ]}$} & & \colhead{${\scriptstyle \left [\frac{10^{-23}}{\sqrt{\Hz}} \right ]}$} & \colhead{${\scriptstyle \left [ \frac{1}{\sqrt{\Hz}} \right ] }$} & \colhead{${\scriptstyle \left [10^{-25} \right ]}$} & & \colhead{${\scriptstyle \left [\frac{10^{-23}}{\sqrt{\Hz}} \right ]}$} & \colhead{${\scriptstyle \left [ \frac{1}{\sqrt{\Hz}} \right ] }$} & \colhead{${\scriptstyle \left [10^{-25} \right ]}$} & & \colhead{${\scriptstyle \left [\frac{10^{-23}}{\sqrt{\Hz}} \right ]}$} & \colhead{${\scriptstyle \left [ \frac{1}{\sqrt{\Hz}} \right ] }$} 
} 
\startdata 
\TBstrut 86.0 & 1.39 & 2.27 & $1.22_{-0.077}^{+0.080}$ & $0.26$ & $1.01$ & $83.1$ & $1.22_{-0.078}^{+0.081}$ & $0.26$ & $0.87$ & $71.5$ & $1.09_{-0.070}^{+0.070}$ & $0.23$ & $0.82$ & $75.4$ \\  
\TBstrut 86.5 & 1.23 & 2.42 & $1.16_{-0.071}^{+0.072}$ & $0.24$ & $1.04$ & $89.7$ & $1.21_{-0.075}^{+0.077}$ & $0.26$ & $0.86$ & $71.2$ & $1.11_{-0.071}^{+0.070}$ & $0.23$ & $0.82$ & $73.9$ \\  
\TBstrut 87.0 & 1.10 & 2.60 & $1.24_{-0.077}^{+0.079}$ & $0.26$ & $1.10$ & $88.4$ & $1.21_{-0.075}^{+0.077}$ & $0.25$ & $0.86$ & $70.7$ & $1.09_{-0.068}^{+0.068}$ & $0.23$ & $0.82$ & $75.5$ \\  
\TBstrut 87.5 & 1.01 & 2.76 & $1.16_{-0.071}^{+0.072}$ & $0.24$ & $1.07$ & $92.2$ & $1.26_{-0.080}^{+0.083}$ & $0.26$ & $0.86$ & $68.5$ & $1.11_{-0.070}^{+0.070}$ & $0.23$ & $0.83$ & $74.9$ \\  
\TBstrut 88.0 & 0.95 & 2.92 & $1.14_{-0.069}^{+0.069}$ & $0.23$ & $1.01$ & $88.7$ & $1.20_{-0.073}^{+0.074}$ & $0.24$ & $0.86$ & $71.5$ & $1.12_{-0.070}^{+0.070}$ & $0.23$ & $0.84$ & $74.6$ \\  
\TBstrut 88.5 & 0.89 & 3.05 & $1.13_{-0.070}^{+0.071}$ & $0.22$ & $0.99$ & $87.2$ & $1.25_{-0.081}^{+0.085}$ & $0.25$ & $0.85$ & $68.2$ & $1.13_{-0.070}^{+0.071}$ & $0.22$ & $0.83$ & $73.2$ \\  
\TBstrut 89.0 & 0.94 & 3.17 & $1.13_{-0.070}^{+0.070}$ & $0.22$ & $0.97$ & $85.8$ & $1.23_{-0.078}^{+0.080}$ & $0.24$ & $0.85$ & $69.2$ & $1.10_{-0.070}^{+0.070}$ & $0.21$ & $0.83$ & $75.2$ \\  
\TBstrut 89.5 & 0.99 & 3.27 & $1.59_{-0.081}^{+0.081}$ & $0.30$ & $0.99$ & $62.0$ & $1.23_{-0.076}^{+0.079}$ & $0.23$ & $0.86$ & $70.0$ & $1.07_{-0.068}^{+0.067}$ & $0.20$ & $0.83$ & $77.3$ \\  
\TBstrut 90.0 & 1.04 & 3.36 & $1.15_{-0.068}^{+0.069}$ & $0.22$ & $0.96$ & $83.3$ & $1.20_{-0.073}^{+0.074}$ & $0.23$ & $0.85$ & $71.1$ & $1.09_{-0.072}^{+0.072}$ & $0.20$ & $0.83$ & $75.9$ \\  
\TBstrut 90.5 & 1.08 & 3.43 & $1.12_{-0.068}^{+0.069}$ & $0.21$ & $0.95$ & $84.7$ & $1.19_{-0.074}^{+0.076}$ & $0.22$ & $0.85$ & $71.1$ & $1.12_{-0.071}^{+0.071}$ & $0.21$ & $0.82$ & $73.3$ \\  
\TBstrut 91.0 & 1.11 & 3.49 & $1.15_{-0.070}^{+0.071}$ & $0.21$ & $0.95$ & $82.9$ & $1.20_{-0.076}^{+0.078}$ & $0.22$ & $0.85$ & $71.1$ & $1.09_{-0.070}^{+0.069}$ & $0.20$ & $0.82$ & $75.4$ \\  
\TBstrut 91.5 & 1.14 & 3.54 & $1.15_{-0.069}^{+0.070}$ & $0.20$ & $0.96$ & $83.2$ & $1.20_{-0.076}^{+0.079}$ & $0.21$ & $0.86$ & $71.9$ & $1.08_{-0.071}^{+0.071}$ & $0.19$ & $0.83$ & $76.9$ \\  
\TBstrut 92.0 & 1.16 & 3.57 & $1.12_{-0.069}^{+0.069}$ & $0.20$ & $0.95$ & $85.0$ & $1.26_{-0.080}^{+0.083}$ & $0.22$ & $0.86$ & $68.3$ & $1.16_{-0.071}^{+0.071}$ & $0.20$ & $0.83$ & $71.3$ \\  
\TBstrut 92.5 & 1.17 & 3.61 & $1.10_{-0.070}^{+0.070}$ & $0.19$ & $0.94$ & $85.9$ & $1.24_{-0.079}^{+0.082}$ & $0.21$ & $0.85$ & $68.9$ & $1.05_{-0.077}^{+0.071}$ & $0.18$ & $0.82$ & $78.2$ \\  
\TBstrut 93.0 & 1.18 & 3.63 & $1.08_{-0.068}^{+0.068}$ & $0.18$ & $0.93$ & $86.2$ & $1.20_{-0.073}^{+0.075}$ & $0.20$ & $0.85$ & $70.8$ & $1.13_{-0.071}^{+0.071}$ & $0.19$ & $0.82$ & $72.9$ \\  
\TBstrut 93.5 & 1.17 & 3.64 & $1.06_{-0.069}^{+0.067}$ & $0.18$ & $0.92$ & $87.0$ & $1.22_{-0.079}^{+0.082}$ & $0.20$ & $0.85$ & $69.5$ & $1.19_{-0.075}^{+0.076}$ & $0.20$ & $0.86$ & $72.0$ \\  
\TBstrut 94.0 & 1.22 & 3.65 & $1.07_{-0.069}^{+0.069}$ & $0.18$ & $0.91$ & $85.4$ & $1.31_{-0.081}^{+0.083}$ & $0.22$ & $0.84$ & $64.1$ & $1.14_{-0.071}^{+0.072}$ & $0.19$ & $0.85$ & $74.6$ \\  
\TBstrut 94.5 & 1.22 & 3.65 & $1.06_{-0.069}^{+0.067}$ & $0.17$ & $0.90$ & $85.2$ & $1.22_{-0.078}^{+0.081}$ & $0.20$ & $0.83$ & $67.9$ & $1.06_{-0.071}^{+0.068}$ & $0.17$ & $0.80$ & $75.9$ \\  
\TBstrut 95.0 & 1.21 & 3.65 & $1.05_{-0.067}^{+0.068}$ & $0.17$ & $0.89$ & $85.1$ & $1.16_{-0.072}^{+0.072}$ & $0.18$ & $0.82$ & $70.3$ & $1.09_{-0.070}^{+0.069}$ & $0.17$ & $0.79$ & $72.2$ \\  
\TBstrut 95.5 & 1.19 & 3.66 & $1.05_{-0.070}^{+0.068}$ & $0.17$ & $0.89$ & $84.3$ & $1.14_{-0.071}^{+0.072}$ & $0.18$ & $0.81$ & $70.8$ & $1.00_{-0.068}^{+0.066}$ & $0.16$ & $0.78$ & $77.9$ \\  
\TBstrut 96.0 & 1.30 & 3.66 & $1.10_{-0.069}^{+0.069}$ & $0.17$ & $0.88$ & $79.8$ & $1.16_{-0.074}^{+0.075}$ & $0.18$ & $0.80$ & $69.1$ & $0.99_{-0.072}^{+0.068}$ & $0.15$ & $0.77$ & $78.3$ \\  
\TBstrut 96.5 & 1.56 & 2.92 & $1.01_{-0.066}^{+0.065}$ & $0.15$ & $0.87$ & $86.5$ & $1.09_{-0.069}^{+0.068}$ & $0.17$ & $0.80$ & $73.2$ & $1.03_{-0.066}^{+0.065}$ & $0.16$ & $0.77$ & $75.0$ \\  
\TBstrut 97.0 & 1.56 & 2.90 & $1.02_{-0.064}^{+0.064}$ & $0.15$ & $0.87$ & $85.5$ & $1.18_{-0.075}^{+0.076}$ & $0.18$ & $0.80$ & $67.5$ & $1.00_{-0.067}^{+0.065}$ & $0.15$ & $0.77$ & $77.5$ \\  
\enddata 
\end{deluxetable*} 


\end{document}